\documentclass[journal,hidelinks]{IEEEtran}

\ifCLASSINFOpdf
\else
\fi
\usepackage{geometry}

\usepackage{amsmath}
\usepackage{amssymb}
\usepackage{url}
\usepackage{xcolor,graphicx}
\usepackage{soul} 

\usepackage{paralist} 
\usepackage{caption}

\usepackage{color} 
\usepackage{setspace} 
\usepackage{rotating} 
\usepackage[english]{babel}
\usepackage[tworuled,linesnumbered]{algorithm2e}
\usepackage{blindtext}

\usepackage{multicol}

\usepackage{mdframed}
\mdfdefinestyle{MyFrame}{%
    linecolor=gray,
    outerlinewidth=20pt,
    roundcorner=10pt,
    innertopmargin=10pt
    innerbottommargin=20pt
    innerrightmargin=10pt,
    innerleftmargin=10pt,
    backgroundcolor=gray!20!white
   }
 \mdfdefinestyle{MyFrameX}{%
    linecolor=gray,
    outerlinewidth=0pt,
    roundcorner=10pt,
    innertopmargin=5pt
    innerbottommargin=5pt
    innerrightmargin=5pt,
    innerleftmargin=5pt,
    backgroundcolor=gray!15!white
   }
\usepackage[ddmmyyyy,hhmmss]{datetime}
\newcommand{\ignore}[1]{}

\newcommand{\fcc}{face-centered-cubic}

\newcommand{\ttx}{$20\times20$}

\newcommand{\abi}{\emph{ab initio}}
\newcommand{\Abi}{\emph{Ab initio}}

\newcommand{\etal}{\emph{et al.}}

\newcommand{\tab}{{\tablename~}}

\newcommand{\fig}{Fig.~}

\renewcommand{\arraystretch}{1.1}

\newcommand{\myTitle}{\huge Protein preliminaries and structure prediction fundamentals for computer scientists}

\newcommand{\mrashid}{Mahmood A. Rashid}
\newcommand{\fkhatib}{Firas Khatib}
\newcommand{\asattar}{Abdul Sattar}

\newcommand{\mrashidx}{\bf M. Rashid}
\newcommand{\fkhatibx}{\bf F. Khatib}
\newcommand{\asattarx}{\bf A. Sattar}

\newcommand{\mrashidE}{mrashid@umassd.edu}
\newcommand{\fkhatibE}{fkhatib@umassd.edu}
\newcommand{\asattarE}{a.sattar@griffith.edu.au}

\newcommand{\CISUMD}{ Department of Computer and Information Science at the University of Massachusetts Dartmouth, MA 02747, USA}

\newcommand{\IIISGU}{Institute for Integrated and Intelligent Systems at Griffith University, Nathan, QLD 4111, Australia}

\newcommand{\jname}{{An arXiv preprint}}

\begin{document}



%
\title{\myTitle}

\author{
	{\mrashid}, 
	{\fkhatib},
    and {\asattar}
    \vspace{-4ex}
	\thanks{{\mrashidx} is with the {\CISUMD}. Email: {\mrashidE}.} 
	\thanks{{\fkhatibx} is with the {\CISUMD}. Email: {\fkhatibE}.}
	\thanks{{\asattarx} is with the {\IIISGU}. Email: {\asattarE}}
}

\markboth{\jname}
{Rashid \MakeLowercase{\etal}: {\myTitle}}

\maketitle

\begin{abstract}
\textbf{Protein structure prediction is a challenging and unsolved problem in computer science. Proteins are the sequence of amino acids connected together by single peptide bond. The combinations of the twenty primary amino acids are the constituents of all proteins. In-vitro laboratory methods used in this problem are very time-consuming, cost-intensive, and failure-prone. Thus, alternative computational methods come into play. The protein structure prediction problem is to find the three-dimensional native structure of a protein, from its amino acid sequence. The native structure of a protein has the minimum free energy possible and arguably determines the function of the protein. In this study, we present the preliminaries of proteins and their structures, protein structure prediction problem, and protein models. We also give a brief overview on experimental and computational methods used in protein structure prediction. This study will provide a fundamental knowledge to the computer scientists who are intending to pursue their future research on protein structure prediction problem.}
\end{abstract}



\IEEEpeerreviewmaketitle
\setcounter{footnote}{0}
\label{chapBackground}

\section{Protein fundamentals}
Proteins are the principal constituents of the protoplasm of all cells. Proteins essentially consist of amino acids, which are themselves bonded by peptide linkages. In this section, we give an overview of protein preliminaries.

\subsection{Amino acids}
\begin{figure}[h]
\centering
\begin{tabular}{cc}
	{\includegraphics[width=5cm]{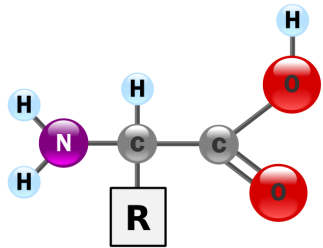}}\\
		{(a)}\\ 
		{\includegraphics[width=5cm]{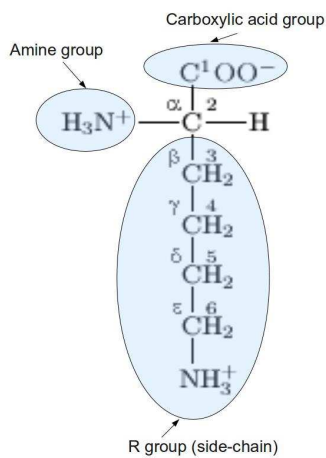}}\\
		{(b)}
\end{tabular}
	\caption{\small (a) The generic structure of an amino acid in its unionized form. (b) Lysine with the carbon atoms in the side-chain.}
	\label{aminoacid}
	\vspace{-3ex}
\end{figure}

Amino acids, also known as protein \emph{monomers} or \emph{residues}, are the molecules containing an amino group, a carboxyl group, and a side-chain.  This side-chain is the only component that varies between different amino acids. Figure \ref{aminoacid} shows the generic structure of an amino acid. An amino acid has the generic formula {H$_2$NCHROOH}. The amino group ({H$_2$N}), the carboxyl group ({COOH}), a hydrogen atom ({H}), and a side-chain (R) are connected to the central carbon denoted by C$_\alpha$. The side chain R is itself an organic subconstituent for all amino acids except Glycine where it stands for a hydrogen atom. The organic side-chain of an amino acid is connected to the C$_\alpha$ by its own carbon atom (denoted by  C$_\beta$). Nevertheless, amino acids are critical to life, and have many functions in metabolism. They are the building blocks of proteins. The twenty primary amino acids that form all the proteins are: Glycine, Alanine, Proline, Valine, Leucine, Isoleucine, Methionine, Phenylalanine, Tyrosine, Tryptophan, Serine, Threonine, Cysteine, Asparagine, Glutamine, Lysine, Histidine, Arginine, Aspartate, and Glutamate.
\vspace{-2ex}

\begin{figure}[!tbh]
	\centering
	\includegraphics[width=.9\columnwidth]{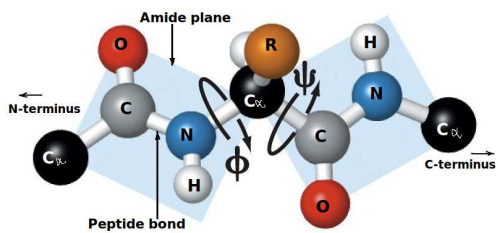}
   \caption{\small Two amino acids are concatenated by forming a peptide bond in between them. $\phi$ and $\psi$ are two bond angles.}
   \label{peptide_bonds}
\end{figure}

\subsection{Proteins}
Proteins are the most important macromolecules in all living organisms. More than half of the dry weight of a cell is made up of proteins of various shapes and sizes { \cite{Voet2004Biochemistry}}. Proteins are basically sequences of amino acids bound into  linear chains. The chain adopts a specific folded three-dimensional (3D) shape and the shape enables the protein to perform specific tasks. Such specific tasks include transporting small molecules (e.g., the hemoglobin transports oxygen in the bloodstream), catalizing biological functions, providing structure to collagen and skin, controlling sense, regulating hormones, and processing emotions { \cite{Pietzsch2003ImpFolding}}.
\begin{figure*}[!tbh]
\centering
	\includegraphics[width=.75\textwidth]{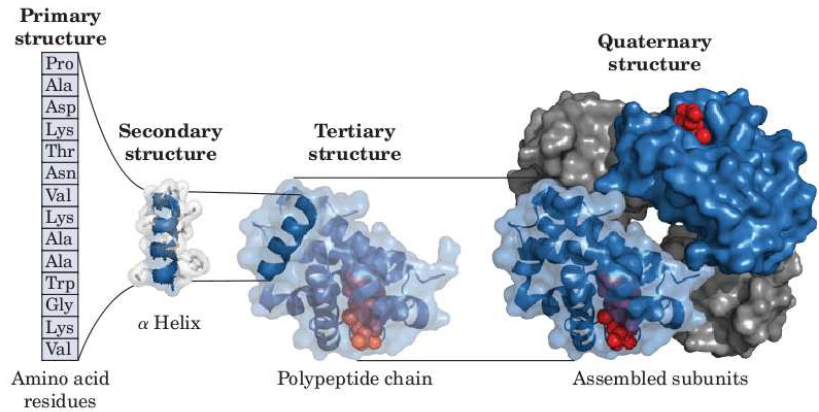}
  		\caption{\small Primary, secondary, tertiary, and quaternary protein structures \cite{NelsonBioChemBook5ThEd}.
  		\label{protein_structure}}
\end{figure*}

In Figure \ref{peptide_bonds}, the two amino acids are bonded together by forming a peptide bond. The carboxyl terminus (known as C-terminus) of one amino acid concatenated with amino terminus (known as N-terminus) of another amino acid to form a peptide bond. This bond creates a rigid amide plane. The number of amino acids concatenated to form a protein determines the length of the protein sequence (or simply the protein length). A protein sequence having $n$ amino acids has $n-1$ peptide bonds and $n-1$ amide planes. The bond angle between N of amine group and C$_{\alpha}$  is denoted by $\phi$ (phi) and the bond angle between C$_{\alpha}$ and C of carboxyl group is denoted by $\psi$ (psi).

\subsection{Protein structures}
As noted before, the proteins fold into 3D structures before performing any task. However, to reach the final 3D structures, the proteins undergo different other structures (Figure \ref{protein_structure}). There are four different level of protein structures. These are: Primary Structure, Secondary Structure, Tertiary Structure, and Quaternary Structure.
	
\subsubsection{Primary structure}	
\begin{figure}[h]
	\centering
	\includegraphics[width=.9\columnwidth]{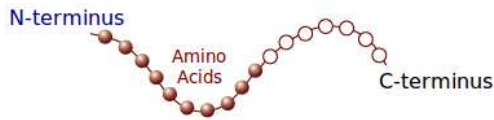}
   \caption{Protein primary structure showing the chain on amino acid sequence.}
   \label{primary_p}
	\end{figure}

	The \emph{Primary structure} as shown in Figure \ref{primary_p}, refers to the amino acid sequence of the polypeptide chain. The primary structure is held together by covalent or peptide bonds, which are formed during the process of protein synthesis or translation.
	
\subsubsection{Secondary structure}	
\begin{figure}[h]
	\centering
	\includegraphics[width=.75\columnwidth]{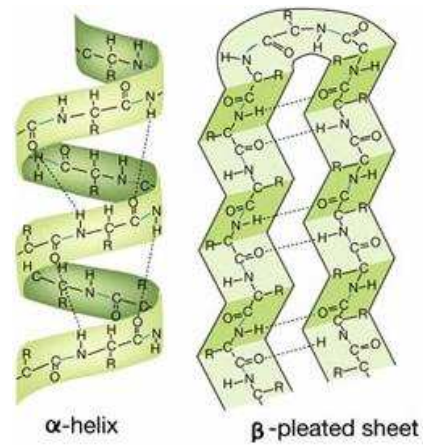}
   \caption{\small Protein secondary structure showing $\alpha$-helix and $\beta$-sheet \cite{Alpha2013Helix}.}
   \label{secondary_p}
	\end{figure}

	The \emph{Secondary Structure} as shown in Figure \ref{secondary_p}, refers to highly regular local sub-structures. In the secondary structure, the alpha helix, the beta sheet, and the random coils are positioned. 
	These secondary structures are defined by patterns of hydrogen bonds between the main-chain peptide groups.  Both the alpha helix and the beta-sheet represent a way of saturating all the hydrogen bond donors and acceptors in the peptide backbone.

\subsubsection{Tertiary structure}	
\begin{figure}[h]
	\centering
	\includegraphics[width=.75\columnwidth]{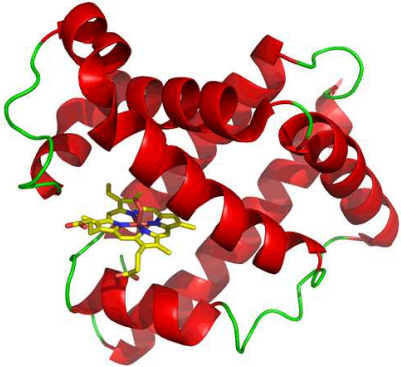}
   \caption{\small Tertiary structure of myoglobin, generated from PDB \cite{PDB2013Intro} file(1mbo.pdb) using PyMOL \cite{PyMOL2013Intro}.}
   \label{tertiary_p}
	\end{figure}

	The \emph{Tertiary structure} as shown in Figure \ref{tertiary_p}, refers to the three-dimensional structure of a single protein molecule. The alpha-helices and beta-sheets are folded into a compact globule. The folding is driven by the non-specific hydrophobic interactions (the burial of hydrophobic residues from water), but the structure is stable \cite{Khan2011BioTech} only when the parts of a protein domain are locked into place by specific tertiary interactions, such as salt bridges, hydrogen bonds, and the tight packing of side chains and disulfide bonds.

\subsubsection{Quaternary structure}	
\begin{figure}[h]
	\centering
	\includegraphics[width=.75\columnwidth]{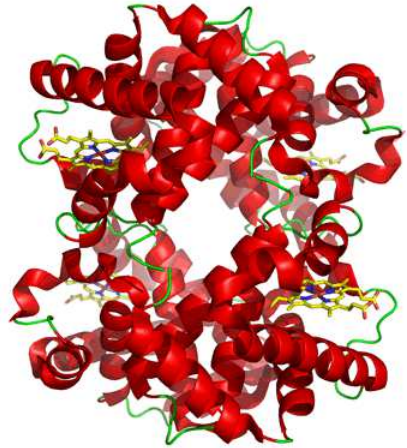}
   \caption{\small Quaternary structure of hemoglobin, generated from PDB \cite{PDB2013Intro} file(1lfq.pdb) using PyMOL \cite{PyMOL2013Intro}.}
   \label{quaternary_p}
   \end{figure}

	The \emph{Quaternary structure} as shown in Figure \ref{quaternary_p}, is a large assembly of several protein molecules or polypeptide chains, usually called subunits in this context. The quaternary structure is stabilized by the same non-covalent interactions and disulfide bonds as the tertiary structure. Complexes of two or more polypeptides (i.e. multiple subunits) are called multimers. Specifically, it would be called a dimer if it contains two subunits; a trimer if it contains three subunits; and a tetramer if it contains four subunits. The subunits are frequently related to one another by symmetry operations, such as a 2-fold axis in a dimer. Multimers made up of identical subunits are referred to with a prefix of ``homo" (e.g. a homotetramer) and those made up of different subunits are referred to with a prefix of ``hetero" (e.g. a heterotetramer, such as the two alpha and two beta chains of hemoglobin). Many proteins do not have the quaternary structure and function as monomers.
	
\subsection{Native structure of protein}
The native structure of a protein is its natural state in the cell, unaltered by heat, chemicals, enzyme action, or the exigencies of extraction. To perform its specific task, proteins need to be folded into their 3D structures and to reach their maximum stable state or globular lowest energy level known as native state  \cite{Pietzsch2003ImpFolding}. A protein can be denatured \cite{Anfinsen1973GovernFolding, Tanford1968Denaturation}, that is, it can have a forced folding deformity, either by adding certain chemicals or by applying heat. It has been experimentally verified that after the denaturing chemical or heat is removed, proteins spontaneously refolded to their native forms. Refolding experiments indicate that the unique native conformation does not depend on the initial state of the chain and is sufficiently stable to be independent of a variety of external factors. 

\subsection{Central dogma of molecular biology}
The central dogma of molecular biology was first articulated by Francis Crick in 1958  \cite{Crick1958CentralDogma} however, re-stated and published in 1970 \cite{crick1970dogma}. The central dogma of molecular biology deals with the detailed residue-by-residue transfer of sequential information. It states that such information cannot be transferred from protein to either protein or nucleic acid.\\

\begin{mdframed}[style=MyFrameX]
	\textsf{\small The central dogma of molecular biology deals with the detailed residue-by-residue transfer of sequential information. It states that such information cannot be transferred from protein to either protein or nucleic acid.}\vspace{1ex}
\end{mdframed}

\begin{figure}[h]
	\centering
	\includegraphics[width=.75\columnwidth]{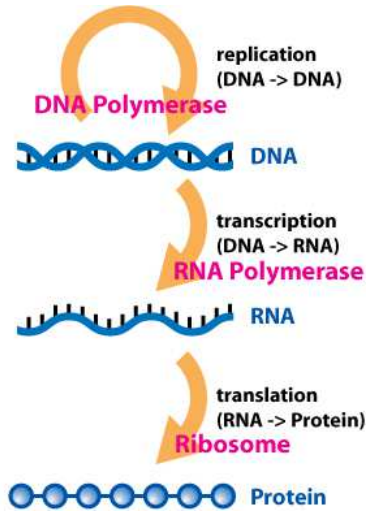}
   \caption{\small Central dogma of molecular biology \cite{Anfinsen1961Dogma}.}
   \label{centraldogma}
\end{figure}

Figure \ref{centraldogma} shows that proteins are irreversible once it has been synthesized. So, it is obvious that the native structure of a protein solely depends on it's amino acid sequence.

\subsection{Molecular driving forces in proteins}
Proteins adopt their three-dimensional shape from a combination of inter macromolecular interactions known as forces. These forces originate from the propensities of the amino acids that make up a protein \cite{NelsonBioChemBook5ThEd}. From {\tab}\ref{aminoacid_properties}, we see that the properties of amino acids can be grouped into different sets---hydrophobic (Column {\sf H}), hydrophilic (Column {\sf P}), neutral (Column {\sf N}), negatively charged (Column {\sf -ve}), and positively charged (Column {\sf +ve}). The main forces that cause the protein folding process can be grouped together into Covalent bonds, Electrostatic forces, Hydrophobicity, van der Waals' interaction, and Hydrogen Sulfide bonds. These forces are briefly illustrated in the following subsections.

	\begin{table}[h]
	\caption[Amino acid properties]{\small The twenty standard amino acids with their properties \cite{NelsonBioChemBook5ThEd}. The columns M, P, N, +ve and -ve stand for hydrophobic, polar, neutral, positive and negative respectively.} 
		\label{aminoacid_properties}
	\centering
		\setlength{\tabcolsep}{2pt}
		\renewcommand{\arraystretch}{1.5}
		\begin{footnotesize}
		
		\begin{tabular}{|r|l|c|c|c|c|c|c|c|}
		\hline	
		&&{\bf Three}&{\bf One}&&&&&\\	
		{\bf SL} & {\bf Amino Acid}& {\bf Alphabet}& {\bf Alphabet}& {\bf H} & {\bf P} & {\bf N} & {\bf +ve} & {\bf -ve} \\
		\hline
		1 & Glycine &Gly&G& $\surd$ & & & &  \\ 
	\hline
	2 & Alanine &Ala&A& $\surd$ & & & &  \\ 
	\hline
	3 & Proline &Pro&P& $\surd$ & & & &  \\ 
	\hline
	4 & Valine &Val&V& $\surd$ & & & &  \\ 
	\hline
	5 & Leucine &Leu&L& $\surd$ & & & &  \\ 
	\hline
	6 & Isoleucine &Ile&I& $\surd$ & & & &  \\ 
	\hline
	7 & Methionine &Met&M& $\surd$ & & & &  \\ 
	\hline
	8 & Phenylalanine &Phe&F& $\surd$ & & & &  \\ 
	\hline
	9 & Tyrosine &Tyr&Y& $\surd$ & & & &  \\ 
	\hline
	10 & Tryptophan &Trp&W& $\surd$ & & & &  \\ 
	\hline
	11 & Serine &Ser&S&  & $\surd$ & $\surd$ & &  \\ 
	\hline
	12 & Threonine &Thr&T&  & $\surd$ & $\surd$ & &  \\ 
	\hline
	13 & Cysteine &Cys&C& & $\surd$ &  $\surd$ & &  \\ 
	\hline
	14 & Asparagine &Asn&N&  & $\surd$ & $\surd$ & &  \\ 
	\hline
	15 & Glutamine &Gln&Q&  & $\surd$ & $\surd$ & &  \\ 
	\hline
	16 & Lysine &Lys&K&  & $\surd$ & & $\surd$ &  \\ 
	\hline
	17 & Histidine &His&H&  & $\surd$ & & $\surd$ &  \\ 
	\hline
	18 & Arginine &Arg&R&  & $\surd$ & & $\surd$ &  \\ 
	\hline
	19 & Aspartate &Asp&D&  & $\surd$ & &  & $\surd$  \\ 
	\hline
	20 & Glutamate &Glu&E&  & $\surd$ & &  & $\surd$ \\ 
		\hline
		\end{tabular}
\end{footnotesize}
	\end{table}

\subsubsection{Covalent bonds}
A covalent bond \cite{Berg2002Bonds} is the strongest type of bond and is the significant one in terms of protein structure. Each atom in an amino acid is covalently bonded together, and the primary structure of a protein has these amino acids within it covalently bonded to one another. Disulfide bridges are another covalent bonds that play an important role in protein structure. The bridges are caused when two amino acids with sulfhydryl groups ($-$SH) come closer during the protein folding process. The sulfur of one cystine  bonds to the other and glues that part of the protein together {\cite{NelsonBioChemBook5ThEd}.
	
\subsubsection{Electrostatics}
Electrostatics \cite{Berg2002Bonds} is the branch of science that deals with the forces arising from stationary or slow-moving electric charges. Electrostatic phenomena arise from the forces that electric charges exert on each other. There are two kinds of interactions in proteins---specific interactions and non-specific interactions. The specific interactions are largely electrostatic interactions such as hydrogen bonds, salt bridges, or ion pairs in the protein structure; whereas, the non-specific interactions are largely hydrophobic and arise due to the burial of the non-polar residues.
	 
\subsubsection*{Ionic bonds}
The ionic bonds \cite{Berg2002Bonds} (a.k.a. \emph{salt bridges}) occur when the positively and the negatively charged amino acids are appeared next to each other within the protein's hydrophobic core. This bond is potent but rare. Because most of the charged amino acids are hydrophilic in nature and, are located on the surface of the proteins. These bonds impact heavily on the three-dimensional structure of a protein as the strength of the ionic bonds can be equal to that of the covalent bonds.
	
\subsubsection*{Hydrogen bonds}
A hydrogen bond is the attractive interaction of a hydrogen atom with an electronegative atom, such as nitrogen, oxygen or fluorine that comes from another molecule or chemical group. The hydrogen must be covalently bonded to another electronegative atom to create the bond. These bonds can occur between molecules (inter-molecularly), or within different parts of a single molecule (intra-molecularly) {\cite{Gold1987HydrogenBond}}. The hydrogen bond (5 to 30 kJ/mole) is stronger than a van der Waals interaction, but weaker than the covalent or the ionic bonds. This type of bond occurs in both inorganic molecules such as water, and organic molecules such as DNA, and Protein.
\begin{figure}[!tbh]
	\begin{tabular}{c}
		\includegraphics[width=.75\columnwidth]{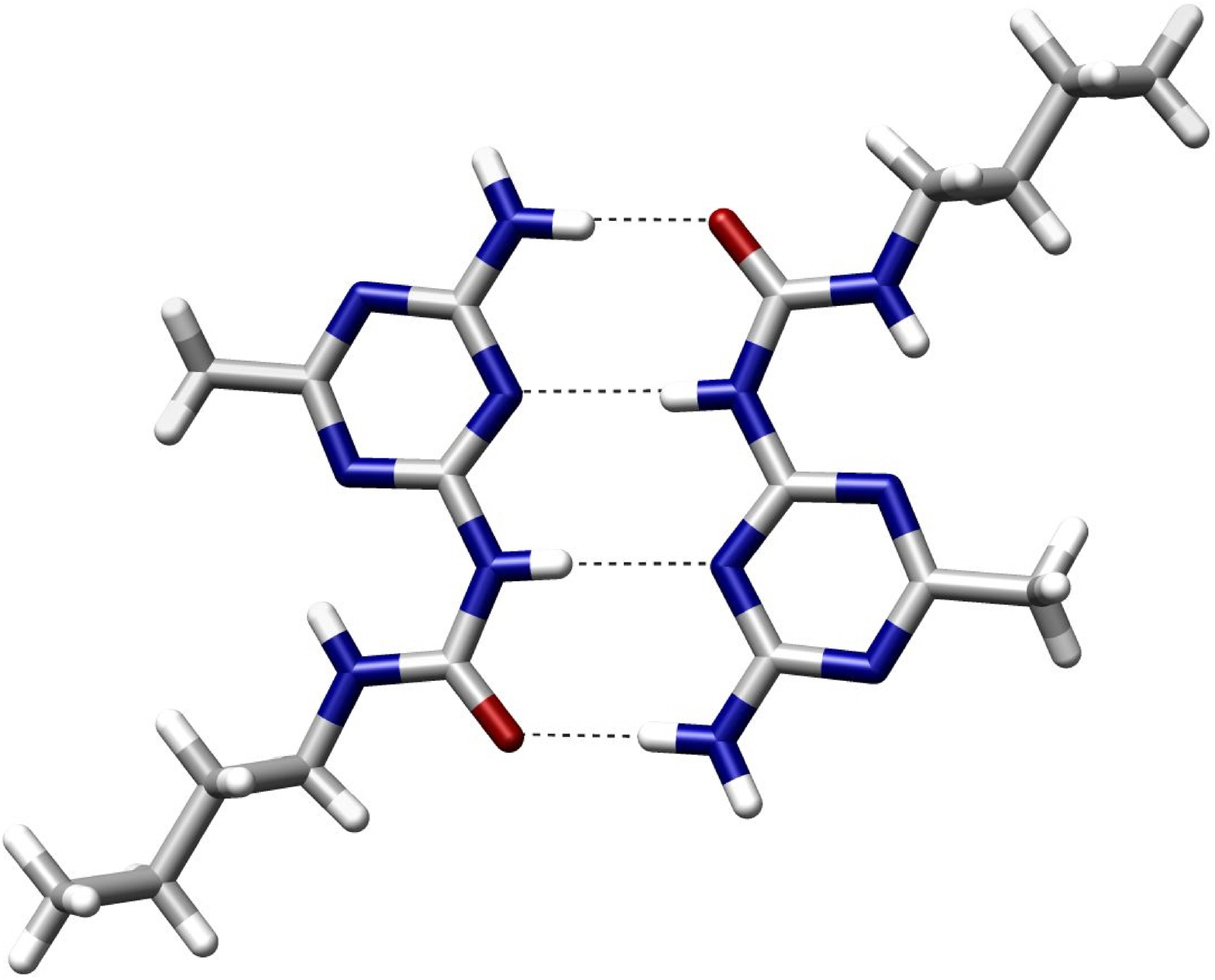}\\
		(a)\\
		\includegraphics[width=.9\columnwidth]{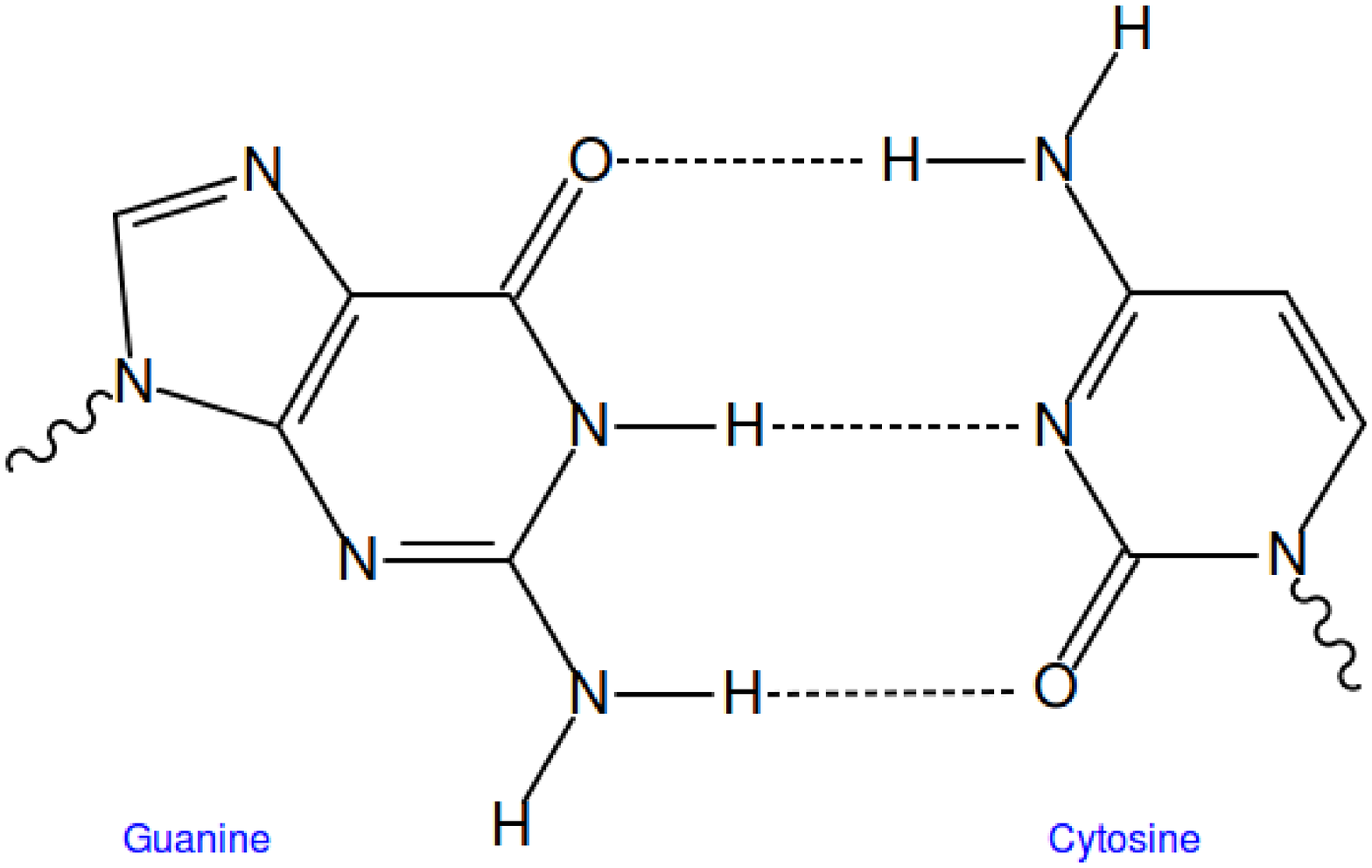}\\
		(b)
	\end{tabular}
   \caption{\small (a) An example of intermolecular hydrogen bonding in a self-assembled dimer complex reported by Beijer and co-workers  \cite{Beijer1998HydrogenBond}. (b) Hydrogen bonding between guanine and cytosine, one of the two types of base pairs in DNA.}
	\end{figure}

Intermolecular hydrogen bonding is responsible for the high boiling point of water (100$^0$C) compared to the other group 16 hydrides that have no hydrogen bonds. Intra-molecular hydrogen bonding is partly responsible for the secondary, tertiary, and quaternary structures of proteins and nucleic acids. 
	
\subsubsection{van der Walls' interactions}
In physical chemistry, the van der Waals force (or van der Waals interaction), named after Dutch scientist Johannes Diderik van der Waals, is the sum of the attractive or repulsive forces between molecules---or between parts of the same molecule---other than those due to the covalent bonds or to the electrostatic interaction of ions with one another or with neutral molecules \cite{Gold1987HydrogenBond}. The term includes: (\emph{i}) force between a permanent dipole and a corresponding induced dipole (Debye force) and (\emph{ii}) force between two instantaneously induced dipoles (London dispersion force).

Sometimes, it is used loosely as a synonym for the totality of intermolecular forces. The van der Waals forces are relatively weak compared to normal chemical bonds, but play a fundamental role in various fields such as molecular chemistry, structural biology, polymer science, nanotechnology, surface science, and condensed matter physics. The van der Waals forces define the chemical character of many organic compounds. 
	
When dealing with macromolecular surfaces making contact, several hundred van der Waals' interactions may be involved between every pair of atoms juxtaposed at optimum separation \cite{Lecture2Interaction}. Crystal structures reveal that the interior of proteins are packed at the same density as solids, implying a high number of close contacts are made in the correctly folded form of the protein. The total van der Waals' interactions of a protein molecule could therefore be the sum of hundreds of kJ/mol.

\begin{table}[h]
\caption[Kyte's and Doolittle's Hydropathy Index]{\small Kyte's and Doolittle's Hydropathy Index { \cite{Kyte1982Amino}}} 
		\label{hydropathy_index}
	\centering
		\setlength{\tabcolsep}{10pt}
		\renewcommand{\arraystretch}{1.5}
		\begin{tabular}{|r|l|c|c|}
		\hline	
		
		SL & {\bf Amino Acid} & {\bf Short} & {\bf Index}\\
		\hline
		1 & Glycine & GLY & -0.4 \\
		\hline
		2 & Alanine & ALA & 1.8  \\
		\hline
		3 & Proline & PRO & 1.6  \\
		\hline
		4 & Valine & VAL & 4.2  \\
		\hline
		5 & Leucine & LEU & 3.8  \\
		\hline
		6 & Isoleucine & ILE & 4.5  \\
		\hline
		7 & Methionine & MET & 1.9  \\
		\hline
		8 & Phenylalanine & PHE & 2.8  \\
		\hline
		9 & Tyrosine & TYR & -1.3  \\
		\hline
		10 & Tryptophan & TRP & -0.9  \\
		\hline
		11 & Serine & SER & -0.8\\ 
		\hline
		12 & Threonine & THR & -0.7 \\ 
		\hline
		13 & Cysteine & CYS & 2.5  \\ 
		\hline
		14 & Asparagine & ASN & -3.5\\ 
		\hline
		15 & Glutamine & GLN & -3.5\\ 
		\hline
		16 & Lysine & LYS &-3.9  \\ 
		\hline
		17 & Histidine & HIS &-3.2 \\ 
		\hline
		18 & Arginine & ARG & -4.5\\ 
		\hline
		19  & Aspartate & ASP & -3.5 \\ 
		\hline
		20 & Glutamate & GLU & -3.5 \\ 
		\hline
		\end{tabular}
		
	\end{table}		
	
	\begin{figure}[h]
	\centering
		\includegraphics[width=.8\columnwidth]{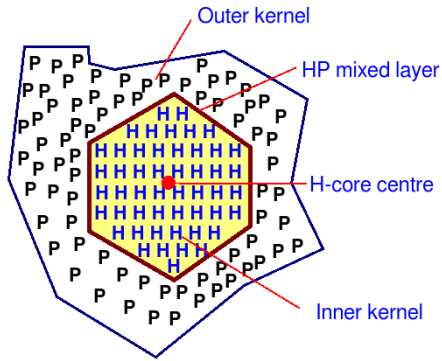}
   \caption{\small The hydrophobic-core in folding kernels.}
   \label{hydrophobic_core}
	\end{figure}

\subsubsection{Hydrophobicity}
The hydrophobicity which is regarded as the major interaction \cite{Dill1985Globular}, stabilizes the tertiary structure of a protein. Hydrophobic amino acids keep themselves away from surrounding water by forming the protein core (See  Figure \ref{hydrophobic_core}) whereas, the hydrophilic (or polar) residues stay outside the protein core, to keep close contact with water. The hydrophobic interaction occurs among the amino acids with hydrophobic side chains while moving towards the center of the protein to minimize the free energy, is caused by the amino acids being exposed to the aqueous (H$_2$O) solvent \cite{Dill1985Globular, NelsonBioChemBook5ThEd}. These interactions lead a protein sequence to forming a hydrophobic core. The hydrophobic intensity of an amino acid can be measured from Kyte's and Doolittle's Hydropathy Index \cite{Kyte1982Amino} as shown in the table \ref{hydropathy_index}.

\section{Protein structure prediction}
 Given a protein's amino acid sequence, the problem is to find a three dimensional structure of the protein such that the total interaction energy amongst the amino acids in the sequence is minimized. Protein structure prediction  (PSP) is the prediction of the secondary, the tertiary, and the quaternary structure of a protein from its primary structure. PSP is one of the most important goals pursued in bioinformatics and theoretical chemistry. t is highly important in drug design and biotechnology. The PSP is computationally a very hard problem \cite{Science2005MuchMore2Know}.
\subsection{Significance of protein structure prediction}
\label{folding_problems}%

The amino acid sequence of a protein determines its structure and the structure determines its mechanism of action. This key paradigm in biochemistry accounts 12 Nobel Prizes in chemistry and physiology or medicine for work in this field from 1956 to 2006 {\cite{Seringhaus2007ChemistryNobel}} which is almost one in four chemistry prizes for structure work. However, in the last decade, fully half Nobel Prizes in chemistry and physiology have dealt with work related to macromolecular structure. This is a significant indication of the importance of the problem and rigorous research works in structural biochemistry OR structural molecular biology.

The function of a protein greatly depends on its folded 3D structure (also known as \emph{native} structure), which has the lowest possible free energy{---the approximation of interaction energies amongst the amino acids in a protein {\cite{Horton1992Energy}}.} {There are, however,} some exceptions such as proteins of \emph{prion domain} {that} have multiple functional structures \cite{King2012Prion}. Many fatal diseases such as prion disease,  Alzheimer's disease, Huntington's disease, Parkinson's disease, diabetes, and cancer are associated with the aggregation of non-functional proteins due to misfolding \cite{Smith2003NatureEd, Dobson2003Misfold, Chiti2006Misfolding, Mathias2013Review}. The 3D structures of proteins are {decidedly} important in rational drug design \cite{Pandi1997Drug, Breda2008Drug}, protein engineering \cite{Kast1997Engg, Berry2001Engg}, and biotechnology \cite{Lilie2003Biotech, Rainer2003Biotech}{; thus,} the protein structure prediction has emerged as an important multi-disciplinary research problem.

\begin{figure}[!tb]
	\centering
	\includegraphics[width=.99\columnwidth]{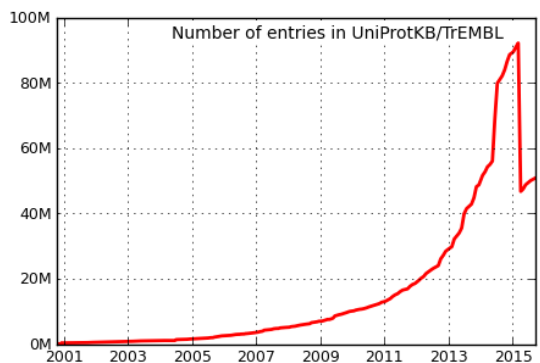}\\
	\caption{\small Yearly growth of sequence data in UniProt database as of release on October 07, 2015.}
	\label{stats_sequence}
\end{figure}

The protein folding problem consists of three closely related puzzles  { \cite{Dill2007FoldingProblem, Dill2008FoldingProblemReview}}: \emph{(i)} What is the folding code? \emph{(ii)} What is the folding mechanism? and \emph{(iii)} Can we predict the native structure of a protein from its amino acid sequence? The general perception has been that the protein folding problem is a grand challenge that will require many supercomputer years to solve  { \cite{Dill2007FoldingProblem}}. Once regarded as a grand challenge, protein folding has seen great progress in recent years.

\begin{figure}[!bth]
	\centering	
	\includegraphics[width=.99\columnwidth]{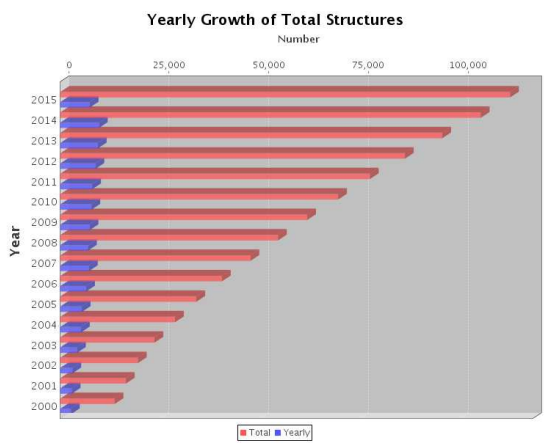}\\
	\caption{\small Yearly growth of total structure in PDB database as of October 07, 2015. The blue bars represent the determined structures for the years whereas, the red bars represent the cumulative numbers.}
	\label{stats_structure}
\end{figure}

%
From numerous research projects, sequences are identifying in much more rapidly (as shown in {\fig}\ref{stats_sequence}) in comparison to determining their structures (as shown in {\fig}\ref{stats_structure}) and thus the number of structure-undetermined proteins are increasing exponentially. According to European Bioinformatics Institute (EBI), the number of identified sequences is $50,825,784$ comprising $16,880,602,444$ amino acids  as of release on October 7, 2015\footnote{EBI website- http://www.ebi.ac.uk/uniprot/TrEMBLstats/}. On the other hand, according to Protein Data Bank (PDB)\footnote{PDB website-  http://www.rcsb.org/pdb/home/home.do}, the number of proteins having their known structures is only $112,722$ as of October 7, 2015. From the statistics it is clear that in PDB, the known structure is $\approx 0.22\%$ of the identified sequences and this is because of the resource intensive and time consuming experimental approaches. Figures \ref{stats_sequence} and  \ref{stats_structure} clearly illustrate the gap between the number of known sequences and the number of known structures. 


\section{Protein structure finding methods}
There are two different approaches for protein structure prediction. They are laboratory based experimental approaches and computational predictive approaches. The  experimental methods are the only reliable methods to determine protein structures. 

\subsection{Experimental methods}
X-ray crystallography \cite{NelsonBioChemBook5ThEd}, Nuclear Magnetic Resonance (NMR) Spectroscopy \cite{Lodish2012, Kurt2001NMR}, Cryo Electron Microscopy (Cryo EM) \cite{CryoEM2013, Bartesaghi2012CryoEM} and Circular Dichroism (CD) Spectroscopy \cite{NelsonBioChemBook5ThEd} are different experimental methods for determining protein structure. Among these, first three methods give 3D information about the protein structures whereas, CD gives one dimensional information about the secondary structure of the protein only. X-ray crystallography, NMR Spectroscopy are two widely used methods where the strengths and weaknesses of one of the two methods fortunately are of those kinds which supplement the holes and gaps in the other method. X-ray crystallography is a very accurate method but proteins need to be crystallized before determining structures. 

\subsubsection{NMR spectroscopy}
Nuclear Magnetic Resonance Spectroscopy most commonly known as NMR spectroscopy, is a research technique that exploits the magnetic properties of certain atomic nuclei to determine physical and chemical properties of atoms or the molecules in which they are contained \cite{Lodish2012,Kurt2001NMR}. It relies on the phenomenon of nuclear magnetic resonance and can provide detailed information about the structure, dynamics, reaction state, and chemical environment of molecules.
			
\subsubsection{X-ray crystallography}
X-ray Crystallography is a method of determining the arrangement of atoms within a crystal, in which a beam of X-rays strikes a crystal and diffracts into many specific directions \cite{Lodish2012}. From the angles and intensities of these diffracted beams, a crystallographer can produce a three-dimensional picture of the density of electrons within the crystal. From this electron density, the mean positions of the atoms in the crystal can be determined, as well as their chemical bonds, their disorder and various other information.


\subsubsection{Limitations of experimental methods}Besides the reliability, the experimental methods have the following  limitations which encourages the computational approaches.

\begin{itemize}
\item {As for NMR Spectroscopy at present, we are limited to less than 200 amino acids or residues in our protein \cite{Lodish2012}. The protein must be soluble and in high concentration. As a rule of thumb, to obtain good NMR data we would need a solution of 30mg/ml.}

\item {X-ray crystallography is a very accurate method but usually as a rule of thumb we should start with milligrams of protein target  \cite{Rupp2004Crystallization}. This is reasonably a large amount. Furthermore, we must be able to crystallize the protein, but not every protein crystallizes and also the structure determined by X-ray is affected by the crystal packing force, which slightly deforms the structure.}

\item {In general, the experimental methods are cost-intensive because of costly resources and are time-consuming as it takes months, year, even couple of years to crystallize the protein to undergo X-ray crystallographic method.}
\end{itemize}

To determine the structures of sequences in rapidly growing sequence-database is highly demanding, and the experimental methods are insufficient to support the process in fast pace. The situation can only be tackled by predicting structures computationally. The  field has been active and open to the researchers for more than a decade to find an effective and reliable computational approach.

\subsection{Computational methods}
In computer science perspective, protein structure prediction is the process of generating a output of 3D structure by taking an amino acid sequence as input. There are three different state-of-the-art predictive methods as described in the following subsections.

\subsubsection{Comparative or homology modeling} 
Homology modeling is based on the reasonable assumption that two homologous proteins will share very similar structures. Because a protein's fold is more evolutionarily conserved than its amino acid sequence, a target sequence can be modelled with reasonable accuracy on a very distantly related template, provided that the relationship between target and template can be discerned through sequence alignment. It has been suggested that the primary bottleneck in comparative modeling arises from difficulties in alignment rather than from errors in structure prediction given a known-good alignment  \cite{Zhang2005PDB}. Unsurprisingly, homology modeling is most accurate when the target and template have similar sequences.

\subsubsection{Protein threading or fold recognition}
Protein threading  \cite{Bowie1991Threading}, also known as fold recognition, is a method of protein modeling which is used to model proteins having the same fold family of proteins with known structures. It differs from the homology modeling as it is used for proteins which do not have their homologous protein  deposited in the Protein Data Bank (PDB). Threading works by using statistical knowledge of the relationship between the structures deposited in the PDB and the sequence of the protein which one wishes to model. 

\begin{figure}[!tb]
\centering
\begin{tabular}{c}
	\includegraphics[width=.8\columnwidth]{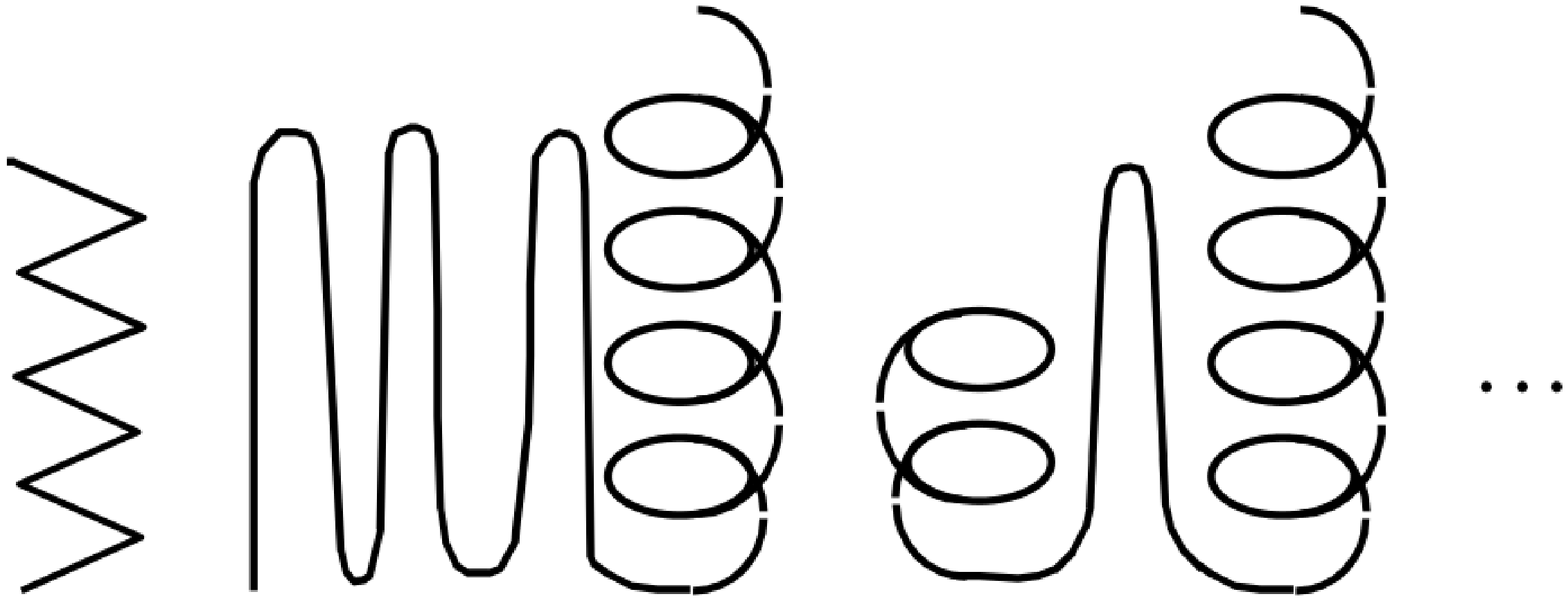}\\
	(a)\\
	\includegraphics[width=.75\columnwidth]{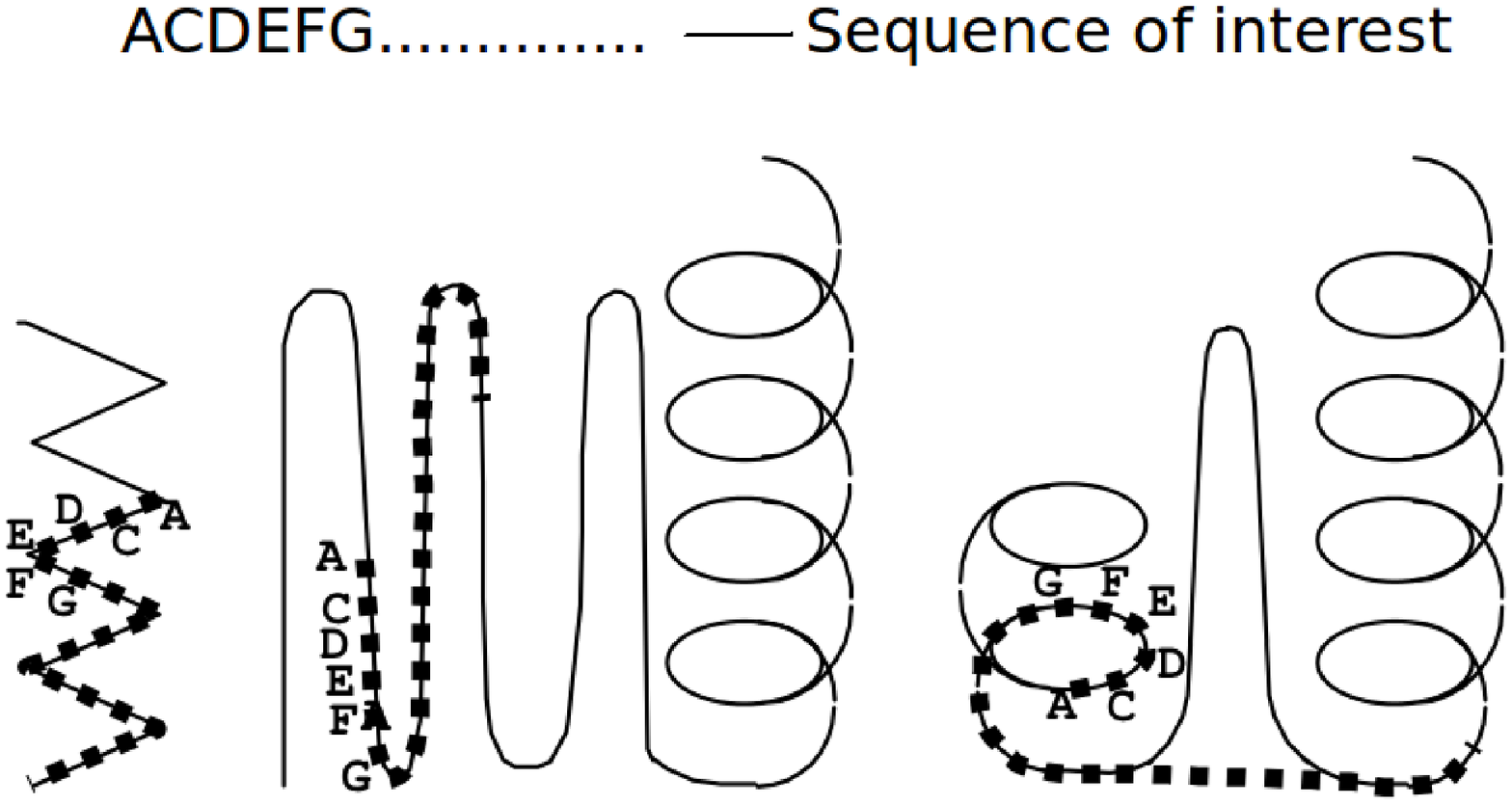}\\
	(b)
\end{tabular}
\caption{\small (a) Structure or template library (500-5000 structures) from PDB. (b) Aligning the sequence of interest with the templates \cite{Torda2005Threading}.}
\label{threading}
\end{figure}

For a threading calculation, there are some elements common to most programs. Firstly, you have a sequence of interest and a library of templates or known structures as shown in Figure \ref{threading}(a). Presumably, these are protein data bank structures and the library contains all known protein folds. Next, one takes the sequence and ``threads" it through each template in the library as shown in Figure \ref{threading}(b). The word threading implies that one drags \cite{Torda2005Threading} the sequence (ACDEFG...) step by step through each location on each template, but really one is searching for the best arrangement of the sequence as measured by some score or quasi-energy function. In the third alignment in Figure \ref{threading}(b), the sequence of interest has been aligned so it skips over part of the template. Finally, all the candidate models with their scores are collected. The best scoring (lowest energy) one is then taken as the structure prediction

\subsubsection{{\Abi} or \emph{de novo} approach}
{\abi} or \emph{de novo} protein modeling methods seek to build three-dimensional protein models `from scratch', i.e., based on physical principles rather than on previously solved structures. There are many possible procedures that either attempt to mimic protein folding or apply some stochastic method to search possible solutions (i.e., global optimization of a suitable energy function). These procedures tend to require vast computational resources, and have thus only been carried out for tiny proteins. To predict protein structures {\abi} for large proteins will require better algorithms and larger computational resources like those afforded by either powerful supercomputers or distributed computing. Although these computational barriers are vast, the potential benefits of structural genomics  make {\abi} structure prediction an active research field.

Levinthal's paradox and Anfinsen's hypothesis are the basis of {\abi} method for protein structure prediction. The idea was originated in 1970 when it was scientifically proven that all information needed to fold a protein resides in its amino acid sequence. It works by conducting a search to find a native conformation, through the conformational search space of protein three-dimensional structures, for a given amino acid sequence of a protein \cite{Simons1999CASP3, Baker2001StructuralGenomics}. 

\begin{figure}[h]
\centering
	\includegraphics[width=.8\columnwidth]{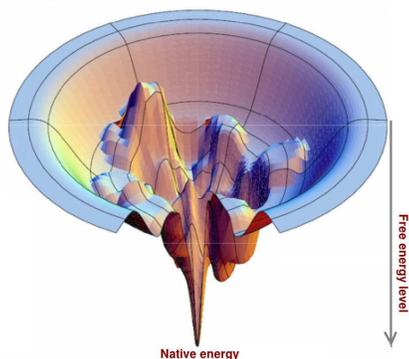}
\caption{\small Energy landscape of the protein folding pathways  \cite{Dill1997Levinthal}}
\label{energy_landscape}
\end{figure}

\begin{itemize}
\item \emph{Levinthal's paradox:}~It is impossible for a protein to go through all of it's possible conformational search space to arrive at its correct native state, since it would take an astronomically large time, while the real proteins take only a few seconds or less to fold. This is known as Levinthal's Paradox \cite{Levinthal1968Pathway}. For example, a simple protein with 100 residues(i.e. amino acids) has 99 peptide bonds and 198 different $\phi$ and $\psi$ angles. if we allow only three possible configurations for these angles, it turns out that the number of possible conformations is $3^{198}$. Therefore if a protein were to attain its correctly folded configuration by sequentially sampling all the possible conformations, it would require a time longer than the age of the universe to arrive at its correct native conformation. This is true even if conformations are sampled at rapid (nanosecond or picosecond) rates. This inspired Levinthal to suggest that there should have some definite pathways of the protein folding 
process.

\item \emph{Anfinsen's thermodynamic hypothesis:} Nobel Prize Laureate Christian Anfinsen's hypothesis  \cite{Anfinsen1973GovernFolding} which states that, at least for small globular proteins, the native structure is determined only by the protein’s amino acid sequence. This amounts to say that, at the environmental conditions (temperature, solvent concentration and composition, etc.) at which folding occurs, the native structure is a unique, stable and kinetically accessible minimum of the free energy.

\item \emph{Energy landscape:} {\fig}\ref{energy_landscape} illustrates the energy landscape for {\abi} protein structure method. At the top of the funnel there is a high concentration of free energy (also known as entropy) and the sequence is in completely unfolded state. The free energy reduces with the progress of protein folding and reaches at the bottom with a stable 3D native structure. The native structure has the least amount of free energy. For completely unknown sequences, where no template is found, {\abi} method produces better results in comparison to protein threading  or homology modeling.

\item \emph{Challenges of the method:}~The {\abi} method has the following main research challenges:
	\begin{compactenum}
		\item conformational search space is astronomically large.
		\item fails to predict structure in case of larger sequences.
		\item energy function is highly complex.
	\end{compactenum}

\item \emph{Dealing with the challenges:}~Several heuristic based non-deterministic search approaches { \cite{Lesh2005Heuristic}} such as  constraint programming   { \cite{Palu2004Constraint, Backofen2006FastExact}}, evolutionary algorithms { \cite{Unger1993GASimulation, Hoque2007PhDThesis}} fragment library based refinement { \cite{Higgs2010Resampling}} have been proposed  to handle first two limitations and the third one also remain unsolved with some progresses { \cite{Backofen1999ExtAlphabet, Dill1995ExactModel, Hoque2009ExtendedHP}}. The results of {\abi} computational method still remains far away from the X-ray crystallographic accuracy.	
\end{itemize}

\begin{figure*}[!tb]
\begin{mdframed}[style=MyFrameX]

\begin{center}
\begin{small}

\begin{equation}
	C_T \approx (\varphi _1 \times \varphi _2 \times ... \times \varphi _{(n-1)})(\psi _1 \times \psi _2 \times ... \times \psi _{(n-1)})(\chi _1 \times \chi _2 \times ... \times \chi _{2n})\;
\label{complexity}
\end{equation}
\vspace{0ex}
\end{small}
\end{center}
\end{mdframed}
\vspace{2ex}
\centering
	\includegraphics[width=.75\textwidth]{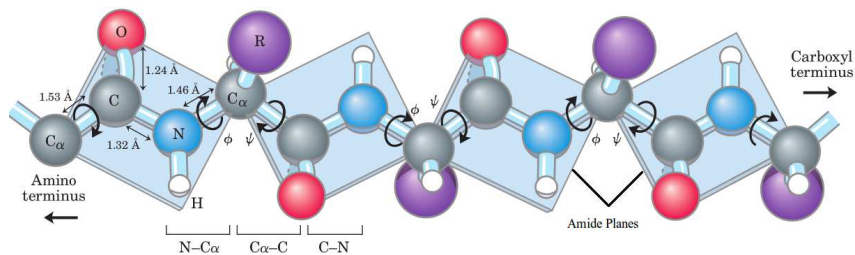}
   \caption[Amino acid's dihedral angles]{A schematic diagram of a amino acid chain \cite{NelsonBioChemBook5ThEd}, showing different angles ($\phi$ and $\psi$) and amide planes.}
   \label{dihedral_angle_nelson}
\end{figure*}

\subsection{Computational complexity of the PSP problem}
\label{computational_complexity}%
A long-standing goal of computational biology has been to devise a computer algorithm that takes, as input, an amino acid sequence and gives, as output, the three-dimensional native structure of a protein. A main motivation is to make drug discovery faster and more efficient by replacing slow expensive structural biology experiments with fast cheap computer simulations. Currently, homology modeling has the speed to compute approximate folds for large fractions of whole genomes. For single-domain globular proteins smaller than about 90 amino acids, web servers can commonly predict native structures often to within about 2-6{\AA} (1 angstrom = 1.0x10$^{-10}$ meters)  of their experimental structures {\cite{Baker2006Macromolecular}}.

The \emph{Science} 2005, named the protein folding problem as one of the 125 biggest unsolved problems in science  \cite{Science2005MuchMore2Know}. The protein folding problem is still unsolved even after more than a decade of rigorous research in protein structure prediction.

\begin{figure}[!tbh]
\centering
\begin{tabular}{c}
	\includegraphics[width=.95\columnwidth]{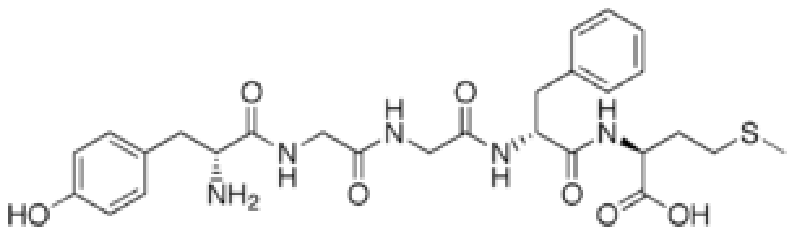}\\
	(a)\\
	\includegraphics[width=.95\columnwidth]{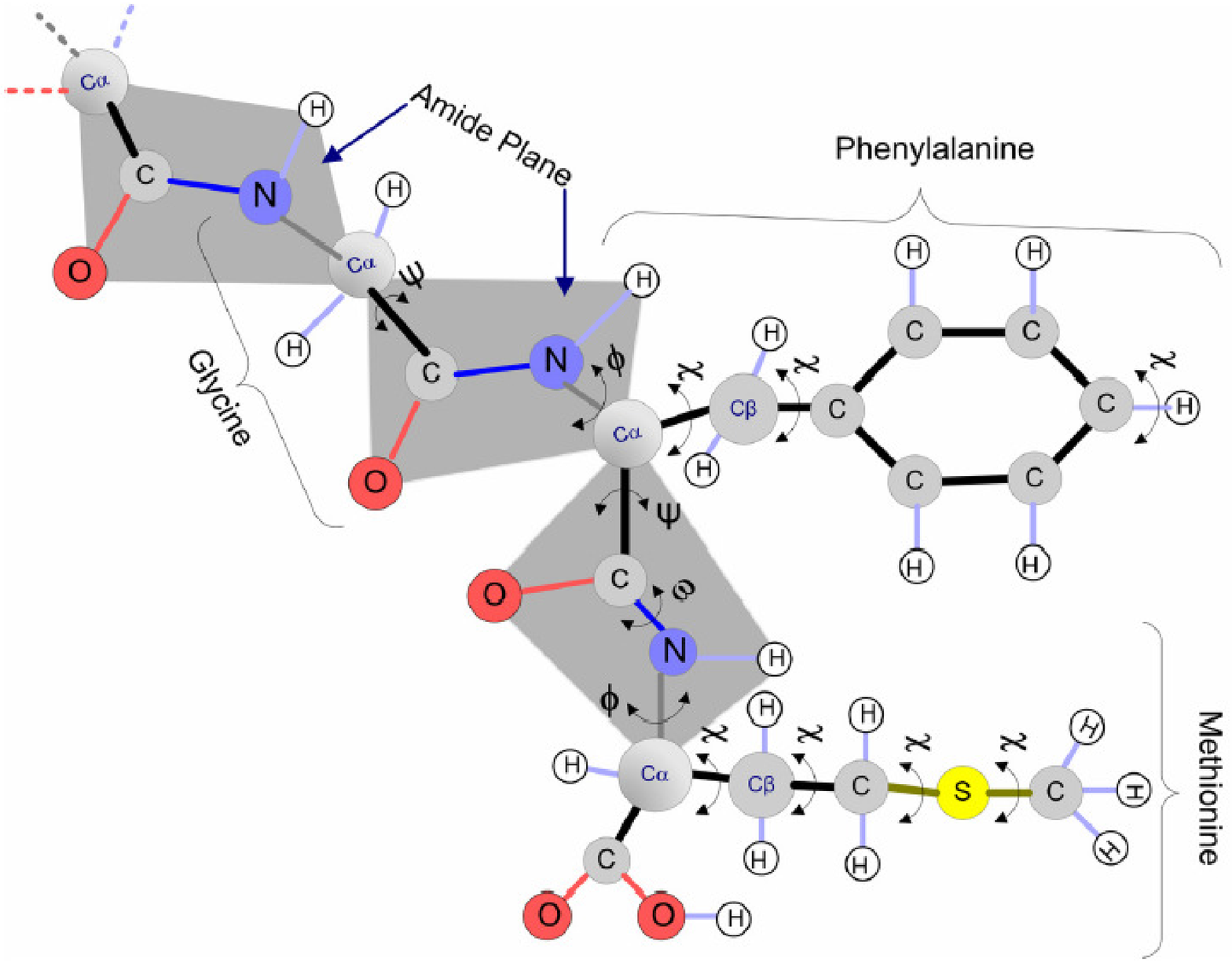}\\
	(b)
\end{tabular}
\caption{\small (a) Met-enkephalin, amino acid sequence \emph{Tyr-Gly-Gly-Phe-Met}. (b) A portion of the [Met]-enkephalin molecule’s concatenated amino acid sequence, \emph{...-Gly-Phe-Met}, showing the formation of the rigid amide plane (shown using shaded plane) and the side chains of the corresponding amino acids. The mobility of the sequence is mostly due to the angles, indicated by $\phi$ and $\psi$ over the connection between N-C$_{\alpha}$ and C$_{\alpha}$-C. The side chain torsion angle is shown by $\chi$  \cite{Hoque2007PhDThesis}. 
}
\label{dihedral_angles}
\end{figure}

From 20 different amino acids, any two can form a peptide bond by joining together, resulting in an amide plane as shown in Figure  \ref{dihedral_angle_nelson}. Both the formation of peptide bonds and the properties of the amide plane are very important in providing specific shape to a specific polypeptide chain formed from the amino acid concatenation. In Figure \ref{dihedral_angles}, three bonds separate sequential ${\alpha}$ carbons in a polypeptide chain has been shown. The N-C$_{\alpha}$ and  C$_{\alpha}$-C bonds can rotate, with bond angles designated $\phi$ and $\psi$, respectively. The peptide C-N bond is not free to rotate and rigid on amide plane as shown in Figure \ref{dihedral_angle_nelson}. In the conformation shown, $\phi$ and $\psi$ have a freedom of 180$^0$ (or -180$^0$). As one looks out from the ${\alpha}$ carbon, the $\phi$ and $\psi$ angles increase as the carbonyl or amide nitrogen (respectively) rotates clockwise. Other single bonds in the backbone may also be rotationally hindered, depending on the size and charge of the R groups. The side chains have an
additional degree of freedom around their torsion angles $\chi$ (See Figure \ref{dihedral_angles}).

To estimate the complexity of the conformation search, let us now examine the crucial question of its feasibility by using all possible combinations of the shape parameters (e.g., dihedral and torsion angles) to determine the optimal conformation through exhaustive enumerations. In fact, we are also interested in determining how many conformations are possible. Assuming that there are n residues in a particular sequence, the total number of conformations ( C$_T$ ) can be expressed as:

where $\varphi _i$ and $\psi _i$ are the two dihedral angles of the $i$th residue and $\chi$ is the torsion angle. Each of the amino acids can have large (depending on the length of the side chain) degrees of freedom for the torsion angles (See Figure \ref{dihedral_angles}). In Equation (\ref{complexity}), on an average two torsion angles ( $\chi _{2i-1}$ and $\chi _{2i}$) are assumed for the $i$th amino acid. However, in practice, for sterical disallowance due to the shape and size of the atoms and also due to their positioning, some reduction in the degree of freedom is possible. The Ramachandran plot  { \cite{Ramachandran1963Chain}} provides a way to visualize dihedral angles $\varphi$ against $\psi$ of amino acid residues in the protein structure, with admissible ranges of these angles in the midst of spatial constraints that are due to steric clashes. Also, since the involved angles can have infinite degrees of freedom, let us consider an arbitrarily chosen discrete finite set of values for each of the angles around $360^0$ for all $i$. This discretization would help measure the immense computational complexity involved.

Interestingly, even if the Equation (\ref{complexity}) is simplified significantly further, its complexity still remains too high. To illustrate, consider restricting each of the amino acid angles $\varphi$ ,$\psi$ and $\chi$ to have three degrees of freedom, that is, to have three possible values. Even with such a massive simplification, considering for instance a 50 residue-long protein sequence, the number of possible conformations is $\approx 3^{(3 \times 50)}$. Thus the search space remains astronomically large, and amongst this large number of conformations, only one conformation assumed to exist which is the native state.  
From the perspective of computational time complexity, assuming a computer can search, typically 200 of these conformations per second, this search would take $\approx 5.8661^{61}$ years to explore all of the possible conformations. Along with the conformational search complexity, there are in reality other forces such as hydrophobic force, hydrogen bonding and electrostatic forces, together with Van der Waals interactions, disulfide bridge and solvation energies, that all serve to influence the final 3D conformation.

\section{Simplified protein models}

To handle the complexities, protein structures are divided into two models: high resolution or real model and low resolution or simplified model. In simplified models the amino acids of a protein are mapped on to a lattice following a self-avoiding-walk, assuming that amino acid bonds are equal in size, and amino acids are never overlapped. 

Electron microscopists have demonstrated \cite{Samudrala1999LowResolution} that low resolution models can yield valuable insights about the function of a protein. Given the large number of sequences being determined and the relatively slow progress of protein structure prediction methods, low resolution models generated by current approaches can be used to elucidate details about structures and functions of the proteins whose atomic structures have not been determined experimentally. A low resolution model, such as an HP based lattice model, is generally used at the first level in PSP investigations. Apart from reducing the complexities, the low resolution model aids in providing a valuable theoretical insight, which is otherwise often very hard to extract in the high resolution model that requires enormously huge computational overhead \cite{Hoque2009Review}. Various current approaches  \cite{Chivian2003Ab, Bonneau2001Progress} involving detail modeling have done so, following a hierarchical paradigm with initial low resolution conformational searching or sampling followed by further investigations with more detailed models. Simplified models can also be off-lattice. Lattice and off-lattice models are discussed in details in the following sub-sections.

%
%

\begin{figure}[!tbh]
	\centering
	\begin{small}
	\begin{tabular}{c}
	\includegraphics[width=.5\columnwidth]{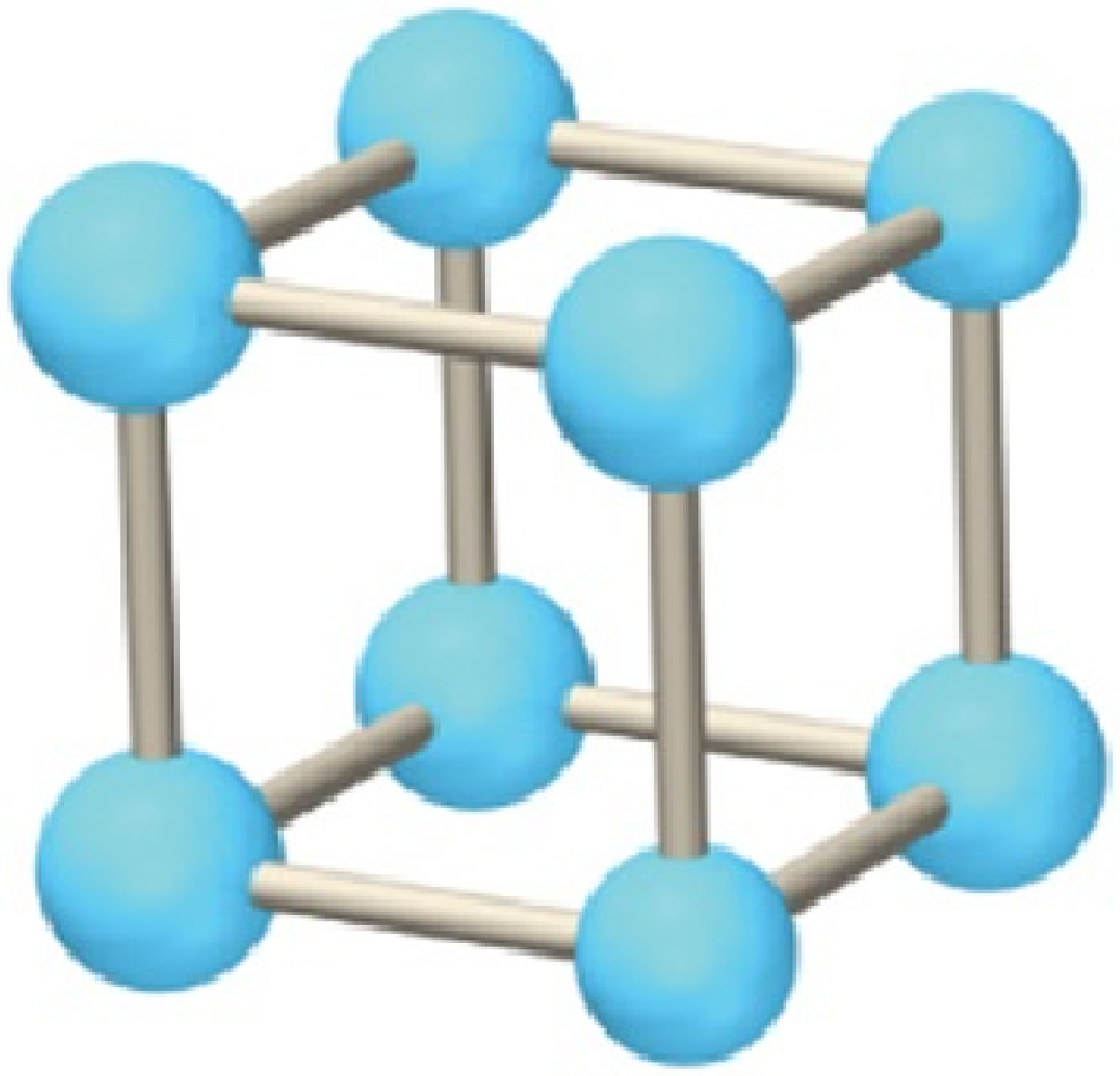}\\
	 {(a) 3D cubic lattice}\\
	\includegraphics[width=.5\columnwidth]{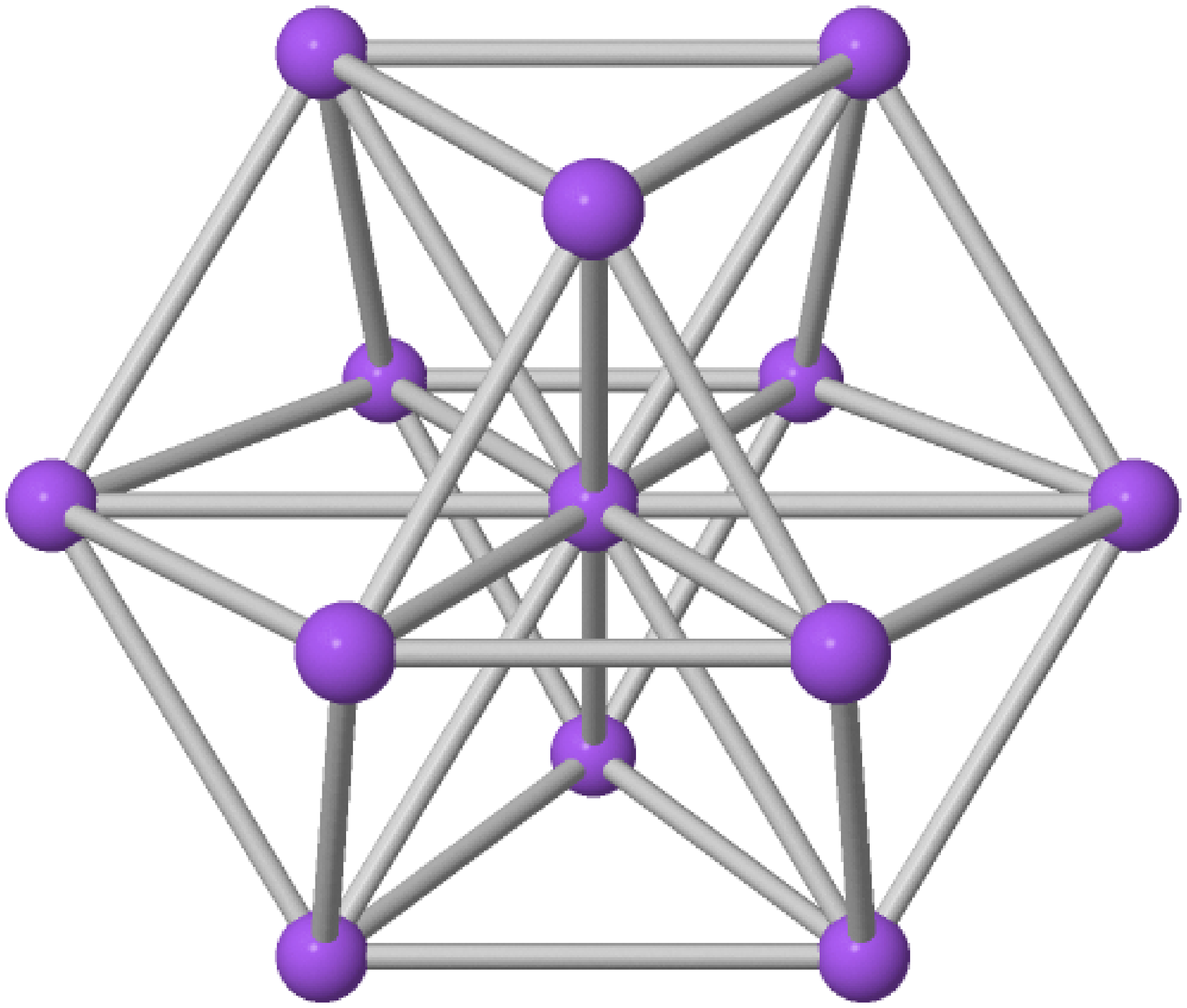}\\
	{(b) 3D HCP lattice}\\
	\includegraphics[width=.5\columnwidth]{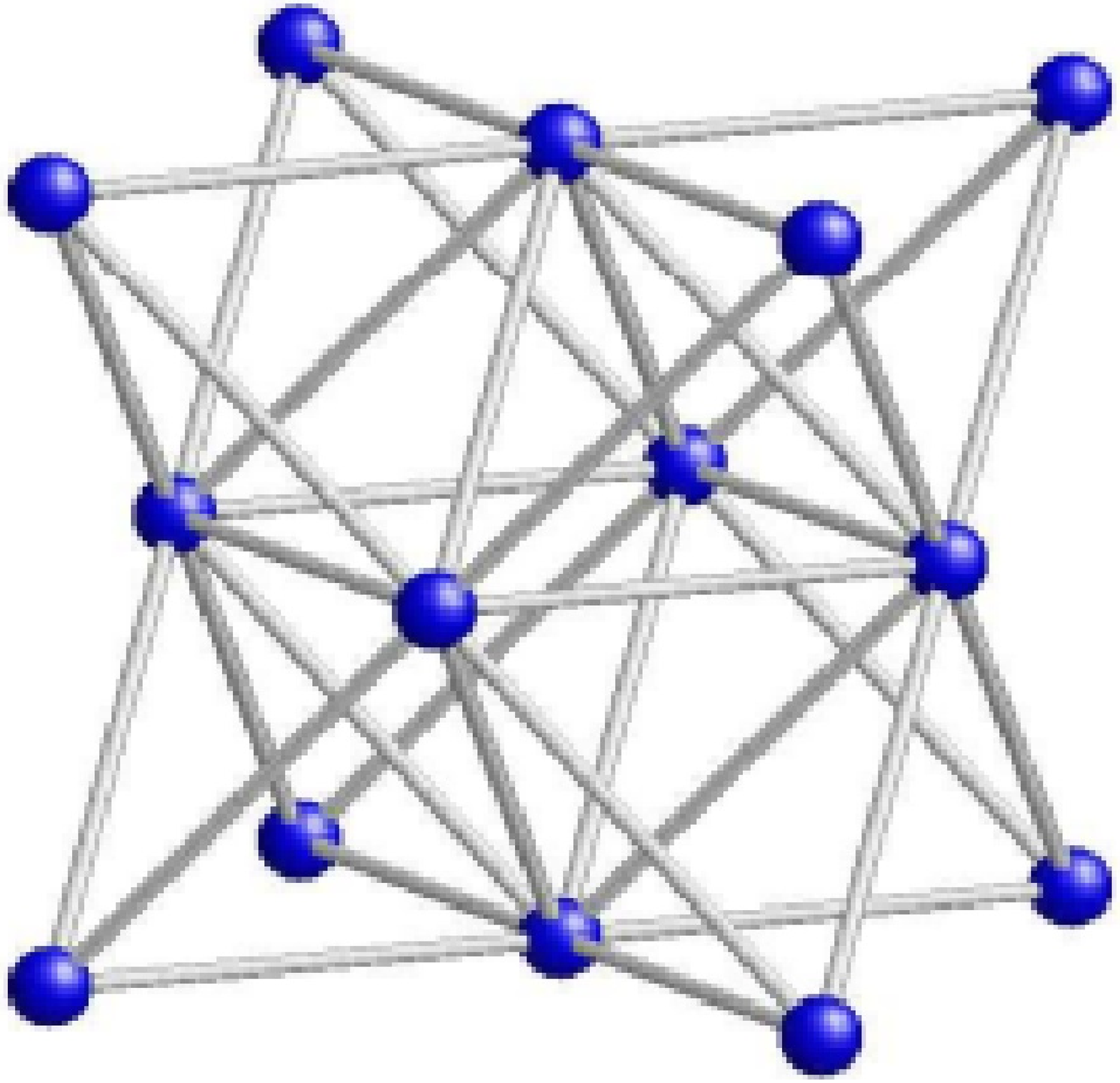}\\
	{(c) 3D FCC lattice}
	\end{tabular}
	\end{small}
	\caption[Popular 3-dimensional lattice models]{\small The three most popular 3-dimensional (3D) lattice models.}
	\label{lattice_3d}
\end{figure}

\subsection{On-lattice models}
A lattice model is a physical model that is defined on a lattice, as opposed to the continuum of space. Currently, lattice models are quite popular in computational physics and computational biology, as the discretization of any continuum model automatically turns it into a lattice model. In simplified PSP model, it assumes that all monomer will stay in  lattice points, creating sequential pathway, for  avoiding overlapping at any point, and the fitness of the structure has been determined from the topological neighbors (TN). Topological neighbors of a lattice node are the nodes that have unit lattice distance away around it. For structure prediction, several standard lattice models (as shown in Figure \ref{lattice_3d}) are widely used for conformation mapping.

\subsubsection{3D cubic lattice}
The cubic lattice (as shown in {\fig}\ref{lattice_3d}a) is the simplest amongst the 3D lattice models. For any lattice point ($x,y,z$) on cubic lattice space, the values of $x$, $y$, and $z$ are always integer numbers. The basis vectors of a cubic lattice are:

\begin{mdframed}[style=MyFrameX]

\begin{center}
\begin{footnotesize}
\setlength{\tabcolsep}{8pt}
\begin{tabular}{l l l}
	$\vec{A}=(1,0,0)$ & $\vec{B}=(-1,0,0)$ & $\vec{C}=(0,1,0)$\\
	$\vec{D}=(0,-1,0)$ & $\vec{E}=(0,0,1)$ & $\vec{F}=(0,0,-1)$\\
\end{tabular}
\end{footnotesize}
\end{center}
\end{mdframed}

\subsubsection{The hexagonal close pack lattice}
The hexagonal close-packed (HCP) lattice (as shown in {\fig}\ref{lattice_3d}b), also known as cuboctahedron, was used in \cite{Hoque2007FCC3D}. In HCP, each lattice point has $12$ neighbors that correspond to $12$ basis vertices with real-numbered coordinates due to the existence of $\sqrt{3}$, which causes the loss of structural precision for PSP. In HCP, each lattice point  has $12$ neighbors with $12$ basis vectors. The basis vectors of a HCP lattice are: %

\begin{mdframed}[style=MyFrameX]

\begin{center}
\begin{footnotesize}
\setlength{\tabcolsep}{4pt}
\begin{tabular}{ll}
$\vec{A}=(1,0,0)$&$\vec{B}=(1/2,\sqrt{3}/2,0)$\\
$\vec{C}=(-1/2,\sqrt{3}/2,0)$&$\vec{D}=(-1/2,-\sqrt{3}/2,0)$\\
$\vec{E}=(1/2,-\sqrt{3}/2,0)$&$\vec{F}=(-1,0,0)$\\
$\vec{G}=(0,1/2,\sqrt{3}/2)$&$\vec{H}=(-\sqrt{3}/4,-1/4,\sqrt{3}/2)$\\
$\vec{I}=(\sqrt{3}/4,-1/4,\sqrt{3}/2)$&$\vec{J}=(0,-1/2,-\sqrt{3}/2)$\\
$\vec{K}=(\sqrt{3}/4,1/4,-\sqrt{3}/2)$&$\vec{L}=(-\sqrt{3}/4,1/4,-\sqrt{3}/2)$\\
\end{tabular}
\end{footnotesize}
\end{center}
\end{mdframed}

\subsubsection{The {\fcc} lattice}
The {\fcc} (FCC) lattice (as shown in {\fig}\ref{lattice_3d}c) has the highest packing density like HCP lattice \cite{Hales2005Proof}. In FCC lattice, each lattice point has $12$ neighbors (as shown below). The advantages of using the FCC lattice over HCP lattice is that the values of $x$, $y$, and $z$ are always integer numbers.

\begin{mdframed}[style=MyFrameX]
\begin{center}
\begin{footnotesize}
\setlength{\tabcolsep}{5pt}
	\begin{tabular}{lll}
			{$\vec{A}=(1,1,0)$}&{$\vec{B}=(-1,-1,0)$}&{$\vec{C}=(-1,1,0)$}\\
			{$\vec{D}=(1,-1,0)$}&{$\vec{E}=(0,1,1)$}&{$\vec{F}=(0,-1,-1$}\\
			{$\vec{G}=(0,1,-1)$}&{$\vec{H}=(0,-1,1)$}&{$\vec{I}=(-1,0,-1)$}\\
			{$\vec{J}=(1,0,1)$}&{$\vec{K}=(-1,0,1)$}&{$\vec{L}=(1,0,-1)$}
	\end{tabular}
\end{footnotesize}
\end{center}
\end{mdframed}

\begin{figure}[!tbh]
	\centering
	\begin{small}
	\setlength{\tabcolsep}{30pt}
	\begin{tabular}{cc}
	\includegraphics[width=.5\columnwidth]{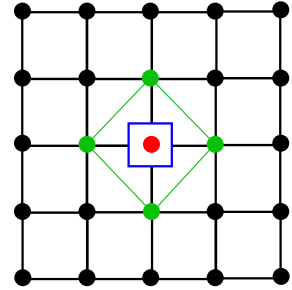}\\
	{(a) Square lattice}\\~\\
	\includegraphics[width=.5\columnwidth]{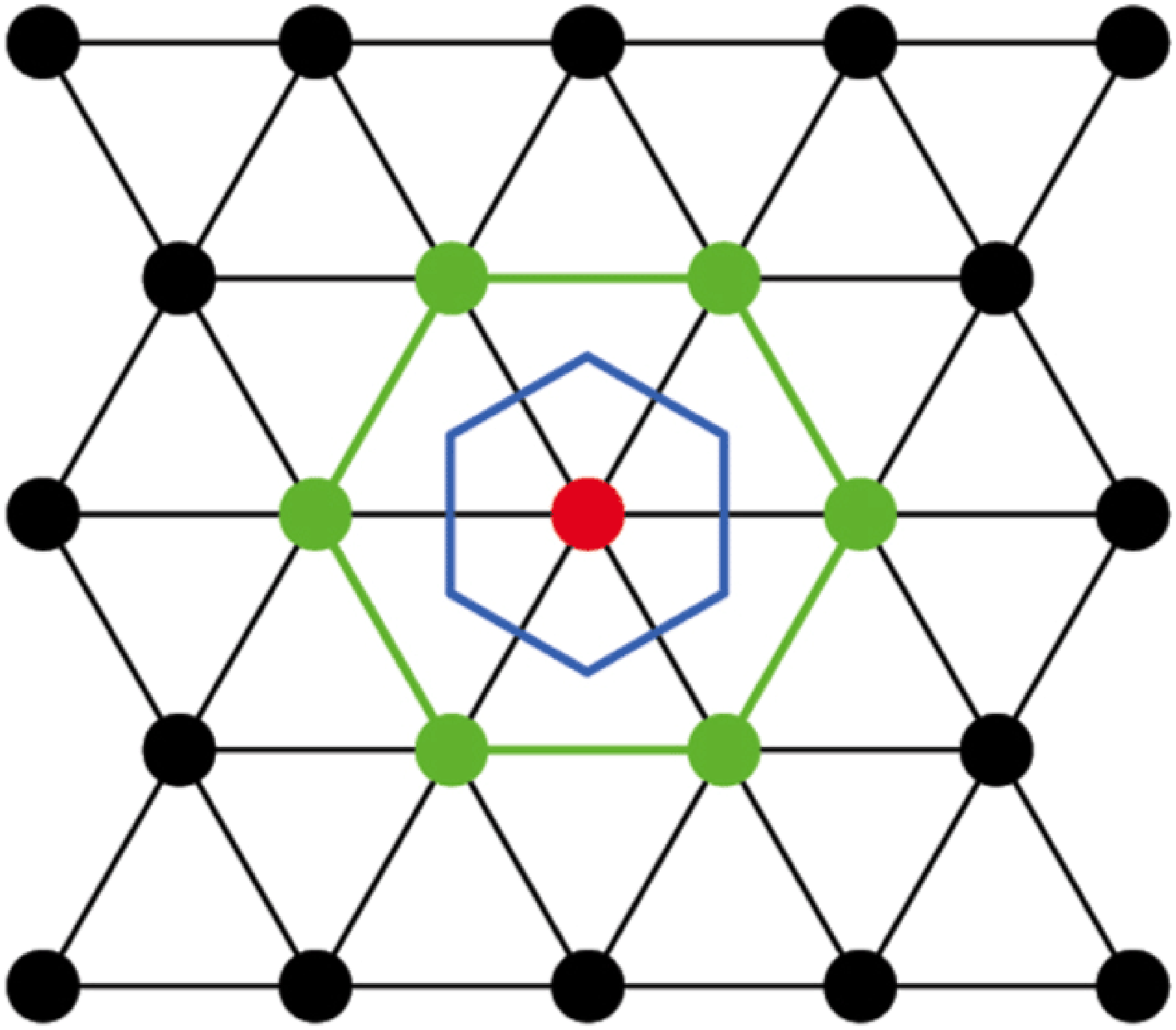}\\
	{(b) Hexagonal lattice \cite{PetkovM09LBM}}
	\end{tabular}
	\end{small}
	\caption{\small The two most popular 2-dimensional (2D) lattice models.}
	\label{lattice_2d}
\end{figure}

\subsubsection{Other lattices}
Besides cubic, HCP, and FCC lattices, there are some 2-dimensional lattices such as  square lattice and triangular lattice as shown in {\fig}\ref{lattice_2d}. The 2D lattice models are used in  \cite{Unger1992GA, Hoque2007PhDThesis, Unger1993GASimulation, Unger1993GAApp} for simplified protein structure prediction. 

For square lattice, each lattice point has $4$ topological neighbors as shown in  {\fig}\ref{lattice_2d}a and the basis vectors are as follows:

\begin{mdframed}[style=MyFrameX]

\begin{center}
\begin{footnotesize}
\setlength{\tabcolsep}{20pt}
\begin{tabular}{ll}
	$\vec{A}=(1,0)$ &$\vec{B}=(-1,0)$\\
	$\vec{C}=(0,1)$ &$\vec{D}=(0,-1)$\\
\end{tabular}
\end{footnotesize}
\end{center}
\end{mdframed}

For hexagonal lattice, each lattice point has $6$ topological neighbors as shown in  {\fig}\ref{lattice_2d}b and the basis vectors are as follows:
\begin{mdframed}[style=MyFrameX]

\begin{center}
\begin{footnotesize}
\setlength{\tabcolsep}{10pt}
\begin{tabular}{ll}
	$\vec{A}=(1,0)$&$\vec{B}=(1/2,\sqrt{3}/2)$\\
	$\vec{C}=(-1/2,\sqrt{3}/2)$ & $\vec{D}=(-1/2,-\sqrt{3}/2)$\\
	$\vec{E}=(1/2,-\sqrt{3}/2)$&$\vec{F}=(-1,0)$\\
\end{tabular}
\end{footnotesize}
\end{center}
\end{mdframed}

\subsection{Off-lattice models}
In off-lattice model \cite{Chen2006OffLattice}, the regular lattice structure is flexible. Both lattice and off-lattice normally start with backbone modeling and then increase the resolution, breaking the residues into further smaller constituents or considering the inclusion of side chains. In the HP tangent sphere model the protein is transformed into a set of tangent spheres of equal radius \cite{Hart1997OffLattice}. Spheres are labelled as either hydrophilic or hydrophobic, and contact between two hydrophobic amino acids is counted when the spheres for those two amino acids are in contact as in Figure \ref{offlattice_tan}.
	
	\begin{figure}[!tbh]
		\centering
		\includegraphics[width=.6\columnwidth]{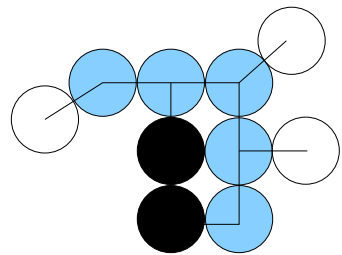}
		\caption{\small The HP tangent sphere side chain model \cite{Hart1997OffLattice} (Off-Lattice). The black lines represent connections between spheres. The light-blue spheres represent the backbone, white spheres represent hydrophilic amino acids and black spheres represent hydrophobic amino acids.}
	\label{offlattice_tan}
	\vspace{-2ex}
	\end{figure}
	
\begin{figure}[h]
		\centering
		\includegraphics[width=.8\columnwidth]{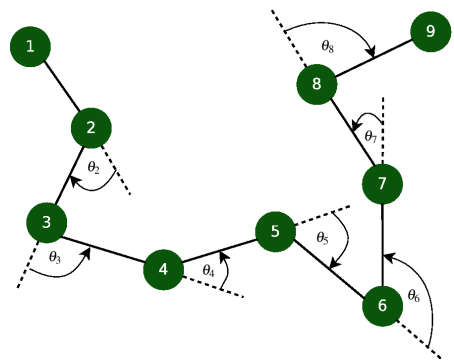}
		\caption{\small A schematic diagram of a generic 9-mer, with serially numbered residues, and backbone bend angles in AB off-lattice model.}
	\label{offlattice_ab_9mer}
	\vspace{-2ex}

	\end{figure}

	AB off-lattice model \cite{Stillinger1993ToyAB} is another significant simple model which also considers two types of monomer. It considers both the sequence dependent Lennard-Jones term that favors the formation of a hydrophobic core and the sequence independent local interactions. The model was first applied in two dimensions  as shown in Figure \ref{offlattice_ab_9mer} and then the model is generalized to three-dimension  \cite{Wang2009ABModel} applying genetic tabu search algorithm.

\subsection{Energy model formulation}
A potential energy function is used to guide the conformational search process or to select a structure from a set of possible candidate structures. All energy functions can be divided into two groups according to whether they are based on statistical analysis of known protein structures (knowledge based) known as  Statistical Effective Energy Function (SEEF) or developed from quantum chemical calculations and incorporation of the water effects (physics based) known as Physical Effective Energy Function (PEEF). It is difficult to use physics based potential functions for proteins since, they are based on the full atomic model and therefore require high computational costs. However, they have the advantage that they exist in real. Another type of potential function is developed through a deductive approach by extracting the parameters of the potential functions from a database of known protein structures. As this approach implicitly incorporates many physical interactions and the extracted potentials, do not necessarily reflect true energies, it is often referred to as ``knowledge-based effective energy function'' \cite{Berrera2003T20}. This approach has quickly gained momentum due to the rapidly growing database of experimentally determined three-dimensional protein structures. These energy functions have the advantage of being simple to construct and easy to use \cite{Lazaridis2000EffEnergy}.

\subsection{HP based energy models}
In protein structure prediction, fitness-function models are very important to evaluate a conformation.  There are several state-of-the-art models such as HP \cite{Dill1985Globular}, HPNX \cite{Bornberg1997Chain}, 1234 \cite{Crippen1991Matrix}, YhHX \cite{Bornberg1997Chain}, and hHPNX \cite{Hoque2009ExtendedHP} models are  used to find out the quality of the predicted structure in terms of some numeric value often known as fitness of the conformation. The details are found in the following subsections. 

\subsubsection{Simple HP model}
\label{sec_hp_model}
The Hydrophobic-Hydrophilic (HP) model was first proposed by Dill and Lau \cite{Dill1985Globular} assuming that hydrophobicity is the major driving force that makes a protein adopt a particular three-dimensional shape. The HP model mimics hydrophobic interactions by modeling a protein structure on lattice using a simplified energy matrix that places hydrophobic residues inside the protein's core and polar residues outside on the protein's surface. 

\begin{figure}[!tbh]
	\centering
	\begin{tabular}{c}
	\includegraphics[width=.5\columnwidth]{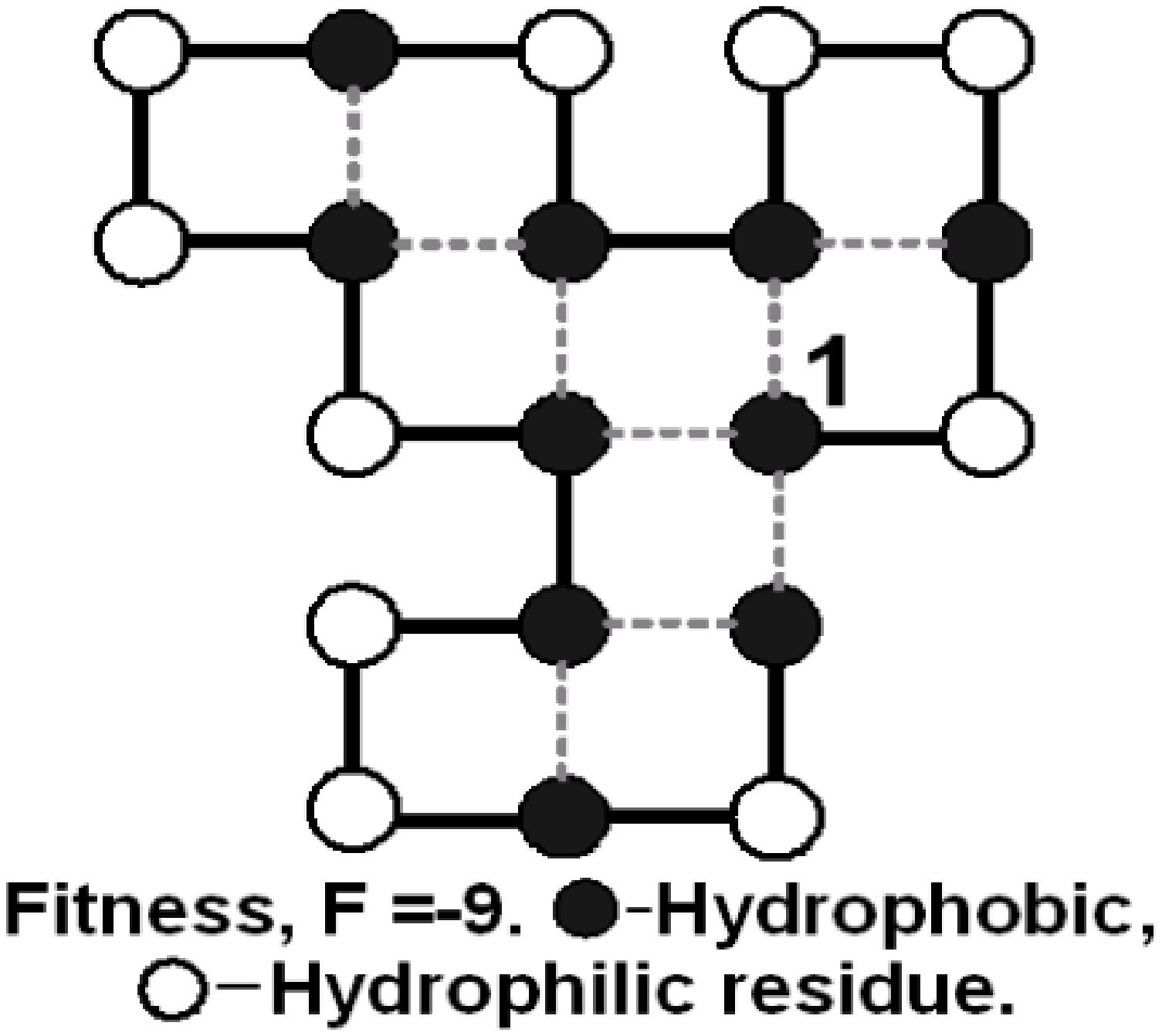}\\
	{(a)}\\
	\includegraphics[width=.5\columnwidth]{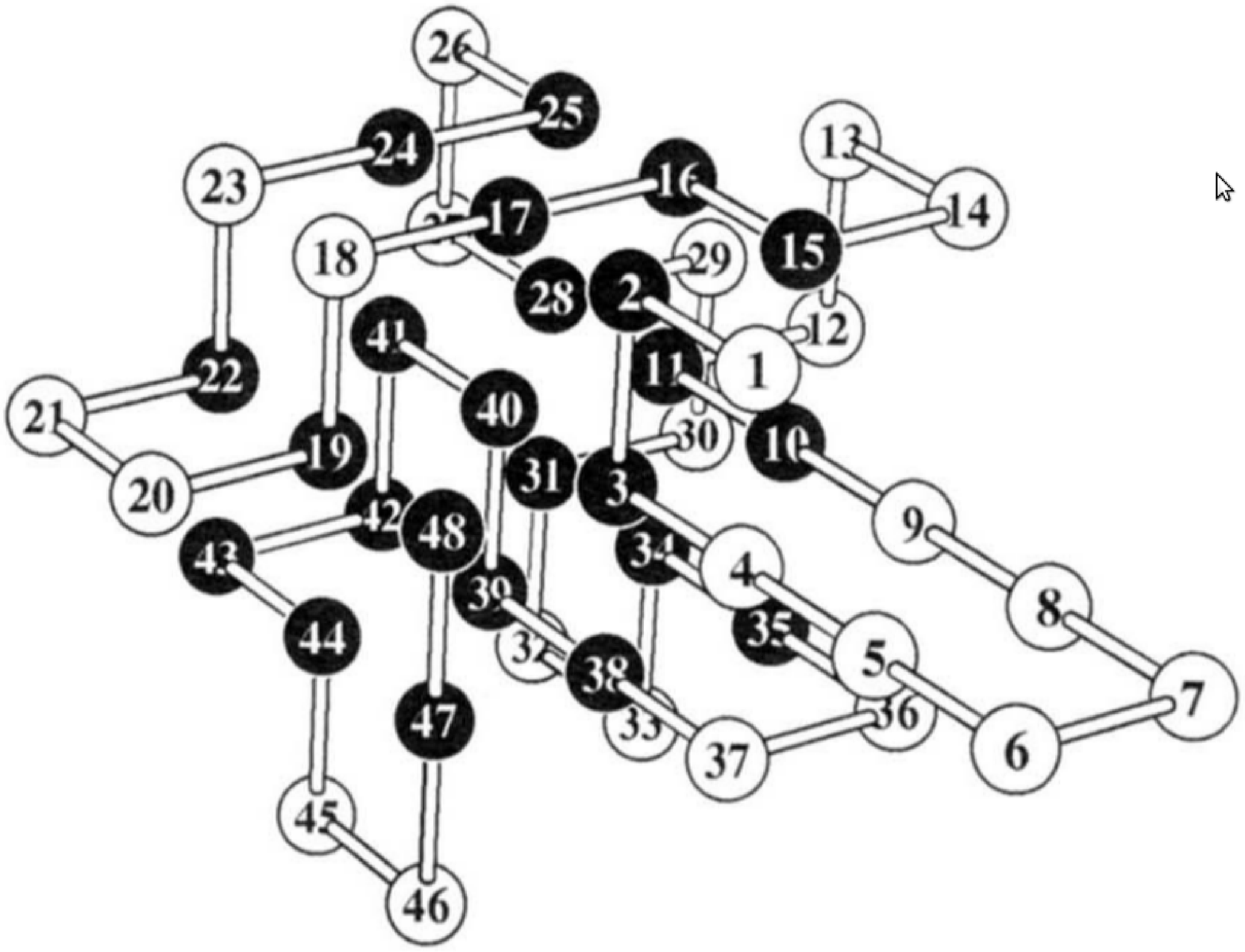} \\ 
	{(b)}\\
	\includegraphics[width=.5\columnwidth]{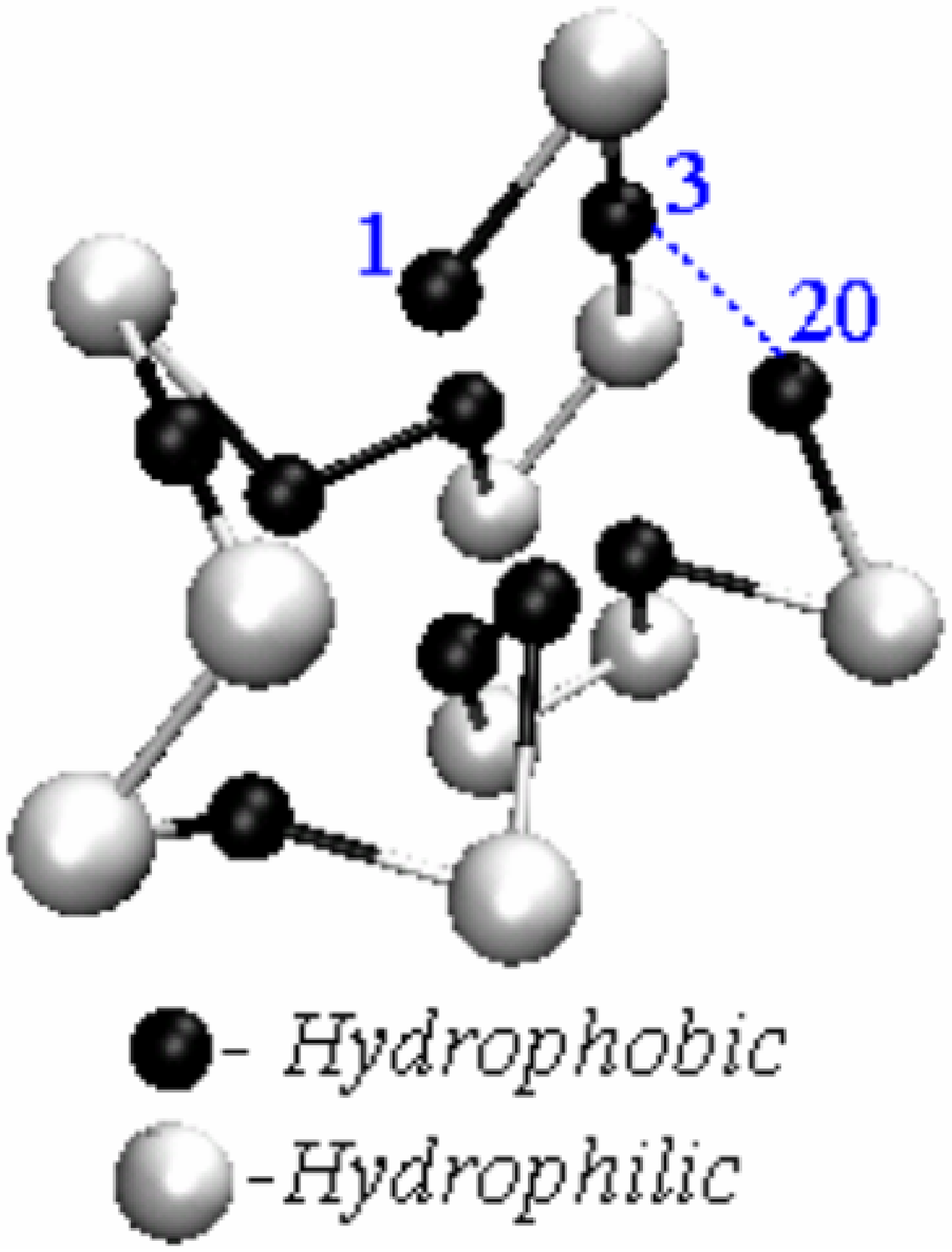}\\
	{(c)}
	\end{tabular}
	\caption{\small (a) A 2D lattice conformation \cite{Hoque2007PhDThesis} (b) a 3D lattice conformation  \cite{Dill1995ExactModel} (c) a 3D FCC lattice conformation  \cite{Hoque2007PhDThesis}}
	\label{hp_lattice_all}
\end{figure}

In simplified lattice models, a particular amino acid sequence the predicted conformation is mapped following a self-avoiding walk (SAW) on 2D square, 3D cube  or 3D FCC cube lattice as shown in Figure \ref{hp_lattice_all}. It simplifies the PSP problem by: (1) only one bead (backbone) or at most two beads (backbone and side-chain) to represent the proteins structure, (2) reducing the conformational search space by constraining the possible conformation onto a regular grid or lattice, (3) bond angles can only have few distinct values, which are restricted due to the structure of the lattice, and (4) the energy function which is heavily simplified \cite{Dill1995ExactModel}.

\begin{table}[h]
\centering
{ 
	\setlength{\tabcolsep}{20pt}
	\renewcommand{\arraystretch}{1.5}
	\begin{tabular}{c}
		\begin{tabular}{|c|c|c|}
			\cline{2-3}
			\multicolumn{1}{c|}{}&{\bf H}&{\bf P}\\
			\hline
			{\bf H}&-1&0\\
			\hline
			{\bf P}&0&0\\
			\hline
		\end{tabular}\\
		(a)\\
		\begin{tabular}{|c|c|c|}
			\cline{2-3}
			\multicolumn{1}{c|}{}&{\bf H}&{\bf P}\\
			\hline
			{\bf H}&-3&-1\\
			\hline
			{\bf P}&-1&0\\
			\hline
		\end{tabular}\\
		(b)\\
		\begin{tabular}{|c|c|c|}
			\cline{2-3}
			\multicolumn{1}{c|}{}&{\bf H}&{\bf P}\\
			\hline
			{\bf H}&-2.5&-1\\
			\hline
			{\bf P}&-1&0\\
			\hline
		\end{tabular}\\
		(c)
	\end{tabular}
	}
	\caption{\small Various energy matrices in HP model. (a) simple energy matrix in HP model \cite{Dill1995ExactModel}, (b) energy matrix by Li et al. 1996  \cite{Li1996Emergence} in HP model (c) energy matrix by Backofen et al. 1999  \cite{Backofen1999ExtAlphabet} in HP model.}
	\label{hp_matrix}
\end{table}

\subsubsection{Crippen's 1234 model}
\label{sec_1234_model}
In \cite{Crippen1991Matrix}, based on the structural observation of a protein data set of 57 protein sequences, Gordon Crippen proposed a potential interaction matrix (shown in {\tab}\ref{1234_matrix}) the amino acids were divided into four different groups. These classifications led to a new four-bead model. The four beads or groups were represented using a single-letter amino acid code as, 1 = $\lbrace$GYHSRNE$\rbrace$, 2 = $\lbrace$AV$\rbrace$, 3 = $\lbrace$LICMF$\rbrace$ and 4 =$\lbrace$PWTKDQ$\rbrace$. Crippen's energy matrix is referred as the \emph{1234} model. He found that the group 2 containing Alanine and Valine had interactions that are consistently different compared to other hydrophobic amino acids. He concluded that the difference in interactions occurred due to their geometrical positions in folded proteins. 
\begin{table}[h]
\centering
{ 
	\setlength{\tabcolsep}{7pt}
	\renewcommand{\arraystretch}{1.5}
	\begin{tabular}{|c|c|c|c|c|}
		\cline{2-5}
		\multicolumn{1}{c|}{}&{\bf 1} & {\bf 2} & {\bf 3}& {\bf 4} \\
		\hline
		{\bf 1} & -0.012 & -0.074 & -0.054 & 0.123\\
		\hline
		{\bf 2} & -0.074 & 0.123 & -0.317 & 0.156\\
		\hline
		{\bf 3} & -0.054 & -0.317 & -0.263 & -0.010\\
		\hline
		{\bf 4} & 0.123 & 0.156 & -0.010 & -0.004\\
		\hline
	\end{tabular}
	}
	\caption{\small Crippen's \emph{1234} potential interaction matrix  \cite{Crippen1991Matrix}}
	\label{1234_matrix}
\end{table}

\subsubsection{HPNX model}
\label{sec_hpnx_model}
Degeneracy problem arises in HP model when more than one conformation contain the same lowest energy or fitness value. To overcome this problem in 1997, Bornberg-Bauer \cite{Bornberg1997Chain} introduced HPNX model, which is different from Crippen's \emph{1234} model. The two beaded HP model \cite{Dill1985Globular} is broken up into 4 beads HPNX. This split is achieved by breaking up the subset of P from the HP model into three subsets based on their electric charges, namely, P-positive, N-negative, and X-neutral. HPNX energy matrix in table \ref{hpnx_matrix} has shown that for H-H contact, P-N contact or N-P contact, fitness is rewarded by some value whereas for any P-P and N-N contact it is penalized.
\begin{table}[h]
\centering
{ 
	\setlength{\tabcolsep}{10pt}
	\renewcommand{\arraystretch}{1.4}
	\begin{tabular}{c}
	\begin{tabular}{|c|c|c|c|c|}
		\cline{2-5}
		\multicolumn{1}{c|}{}&{\bf H} & {\bf P} & {\bf N}& {\bf X} \\
		\hline
		{\bf H} & -4 & 0 & 0 & 0\\
		\hline
		{\bf P} & 0 &  \textcolor{red}{\bf{0}} & -1 & 0\\
		\hline
		{\bf N} & 0 & -1 &  \textcolor{red}{\bf{0}} & 0\\
		\hline
		{\bf X} & 0 & 0 & 0 & 0\\
		\hline
	\end{tabular}\\
	(a)\\~\\
	\begin{tabular}{|c|c|c|c|c|}
		\cline{2-5}
		\multicolumn{1}{c|}{}&{\bf H} & {\bf P} & {\bf N}& {\bf X} \\
		\hline
		{\bf H} & -4 & 0 & 0 & 0\\
		\hline
		{\bf P} & 0 &  \textcolor{blue}{\bf{1}} & -1 & 0\\
		\hline
		{\bf N} & 0 & -1 &  \textcolor{blue}{\bf{1}} & 0\\
		\hline
		{\bf X} & 0 & 0 & 0 & 0\\
		\hline
	\end{tabular}\\
	(b)
	\end{tabular}
	}
	\caption{\small(a) HPNX energy matrix   \cite{Bornberg1997Chain}, (b) HPNX energy matrix   \cite{Backofen1999ExtAlphabet}.}
	\label{hpnx_matrix}
\end{table}

\subsubsection{YhHX model}
\label{sec_yhhx_model}
In \cite{Bornberg1997Chain}, the simplified version of Crippen's \emph{1234} model (as shown in table \ref{1234_matrix}) is proposed by converting the potential matrix from real values into integer values, maintaining the same ratio and referred the new one as YhHX model as shown in table \ref{yhhx_matrix}. Integer values make the model more popular among the researchers  \cite{Buchler2000Surveying}. He found a  value of -2 for interaction between group 2 = $\lbrace$AV$\rbrace$ but, Crippen mentioned that Alanine and Valine repulses each other. Bornberg-Bauer present a corrected matrix mentioning a interaction value between group 2 = $\lbrace$AV$\rbrace$ is +2 rather than -2 because of their repulsion to each other.

\begin{table}[!tbh]
\centering
{ 
	\setlength{\tabcolsep}{10pt}
	\renewcommand{\arraystretch}{1.4}
	\begin{tabular}{c}
	\begin{tabular}{|c|c|c|c|c|}
		\cline{2-5}
		\multicolumn{1}{c|}{}&{\bf Y} & {\bf h} & {\bf H}& {\bf X} \\
		\hline
		{\bf Y}& 0 & -1 & -1 & 2\\
		\hline
		{\bf h} & -1 & \textcolor{red}{\bf{-2}} & -4 & 2\\
		\hline
		{\bf H} & -1 & -4 & -3 & 0\\
		\hline
		{\bf X}  & 2 & 2 & 0 & 0\\
		\hline
		{\bf f$_{q}$} & 10 & 16 & 36 & 28\\
		\hline
	\end{tabular}\\
	(a)\\~\\
	
	\begin{tabular}{|c|c|c|c|c|}
		\cline{2-5}
		\multicolumn{1}{c|}{}&{\bf Y} & {\bf h} & {\bf H}& {\bf X} \\
		\hline
		{\bf Y} & 0 & -1 & -1 & 2\\
		\hline
		{\bf h} & -1 & \textcolor{blue}{\bf{2}} & -4 & 2\\
		\hline
		{\bf H} & -1 & -4 & -3 & 0\\
		\hline
		{\bf X} & 2 & 2 & 0 & 0\\
		\hline
		{\bf f$_{q}$} & 10 & 16 & 36 & 28\\
		\hline
	\end{tabular}\\
	(b)
	\end{tabular}
}
	\caption{\small(a) YhHX matrix: as converted by Bornberg in  \cite{Bornberg1997Chain} from Crippen's matrix and presented in the same order as the Crippen's matrix. (b) Corrected YhHX as it should have been considered in  \cite{Bornberg1997Chain}. Here, $f_{q}$ implies the percentage of occurrence frequencies of amino acid for each of the four groups.}
	\label{yhhx_matrix}
\end{table}

\subsubsection{\emph{hHPNX} model}
\label{sec_hhpnx_model}
In 2008, an extension of HPNX \cite{Bornberg1997Chain} model was proposed by Hoque et al.  \cite{Hoque2009ExtendedHP} and referred as hHPNX model. He performed some experiments again on Crippen's matrix and proved that there was an error in the h-h potential interaction in YhHX  \cite{Bornberg1997Chain}. According to their results Alanine and Valine consistently exhibited repulsion, disproving that the h-h potential interaction value -2 was correct. Instead, Hoque stated that +2 was the correct value, because -2 shows an affinity between Alanine and Valine, when they clearly repel one another.
\begin{table}[!tbh]
\centering
{ 
	\setlength{\tabcolsep}{10pt}
	\renewcommand{\arraystretch}{1.4}
	\begin{tabular}{|c|c|c|c|c|c|}
		\cline{2-6}
		\multicolumn{1}{c|}{}& {\bf h} & {\bf H} & {\bf P} & {\bf N} & {\bf X} \\
		\hline
		{\bf h}  & 2 & -4 & 0 & 0 & 0\\
		\hline
		{\bf H} & -4 & -3 & 0 & 0 & 0\\
		\hline
		{\bf P} & 0 & 0 & 1 & -1 & 0\\
		\hline
		{\bf N} & 0 & 0 & -1 & 1 & 0\\
		\hline
		{\bf X} & 0 & 0 & 0 & 0 & 0\\
		\hline
	\end{tabular}
	}
	\caption{hHPNX energy matrix  \cite{Hoque2009ExtendedHP}}.
	\label{hhpnx_matrix}
\end{table}

To reduce the degeneracy of other HP models and due to the significance of h-h interaction compared to other interactions in the protein dataset, the h has been added to HPNX model and introduced a new hHPNX model and the value of potential interactions depicted as in {\tab}\ref{hhpnx_matrix}.

\subsection{The {\ttx} energy models}

Twenty different amino acids are the primary constituents of proteins. By analyzing crystallized protein structures, Miyazawa and Jernigan \cite{Miyazawa1985T20} statistically deduced a {\ttx} energy matrix (MJ model) considering residue contact propensities.  By calculating empirical contact energies on the basis of information available from selected protein structures and following the quasi-chemical approximation Berrera {\etal} \cite{Berrera2003T20} deduced another  {\ttx} energy matrix.

The total energy $E_{20\times20}$ (as shown in Equation \ref{eqT20}), of a conformation based on the {\ttx} energy matrices becomes the sum of the contributions of all pairs of non-consecutive amino acids of unit lattice distance apart. The contributions  are the empirical energy value between the amino acid pairs  obtained from a given {\ttx} energy matrix.


\begin{mdframed}[style=MyFrameX]
\begin{small}
\begin{equation}
		 E_{20\times20}=\sum_{i<j-1} \mathsf{c}_{ij}.\mathsf{e}_{ij}
		\label{eqT20}
	\end{equation}
\end{small}
\end{mdframed}

Here, $c_{ij} = 1$ if amino acids $i$ and $j$ are non-consecutive neighbors on the lattice, otherwise 0; and $\mathsf{e}_{ij}$ is the empirical energy value between the $i\mathsf{th}$ and $j\mathsf{th}$ amino acid pair.

\section{Summary}
In this study, we present the preliminaries of proteins. In a nutshell, we have discussed amino acids, proteins, protein structures, protein structure determining methods, protein structure prediction methods, and different energy models. We also have illustrated the complexities and significance of the PSP problem. We do believe that this article will provide the fundamental knowledge to the computer scientists who are intending to pursue their future research on protein structure prediction.


\vspace{-1ex}
\begin{small}
	\bibliographystyle{IEEEtran}
	\bibliography{IEEEabrv,psp}

\begin{thebibliography}{10}
\providecommand{\url}[1]{#1}
\csname url@samestyle\endcsname
\providecommand{\newblock}{\relax}
\providecommand{\bibinfo}[2]{#2}
\providecommand{\BIBentrySTDinterwordspacing}{\spaceskip=0pt\relax}
\providecommand{\BIBentryALTinterwordstretchfactor}{4}
\providecommand{\BIBentryALTinterwordspacing}{\spaceskip=\fontdimen2\font plus
\BIBentryALTinterwordstretchfactor\fontdimen3\font minus
  \fontdimen4\font\relax}
\providecommand{\BIBforeignlanguage}[2]{{%
\expandafter\ifx\csname l@#1\endcsname\relax
\typeout{** WARNING: IEEEtran.bst: No hyphenation pattern has been}%
\typeout{** loaded for the language `#1'. Using the pattern for}%
\typeout{** the default language instead.}%
\else
\language=\csname l@#1\endcsname
\fi
#2}}
\providecommand{\BIBdecl}{\relax}
\BIBdecl

\bibitem{Voet2004Biochemistry}
D.~Voet and J.~G. Voet, \emph{{Biochemistry}}, 3rd~ed.\hskip 1em plus 0.5em
  minus 0.4em\relax Wiley: Hoboken, NJ, 2004, vol.~1.

\bibitem{Pietzsch2003ImpFolding}
\BIBentryALTinterwordspacing
J.~Pietzsch, ``{The importance of protein folding},'' \emph{Nature}, 2003.
  [Online]. Available:
  \url{www.nature.com/horizon/proteinfolding/background/importance.html}
\BIBentrySTDinterwordspacing

\bibitem{NelsonBioChemBook5ThEd}
D.~L. Nelson and M.~M. Cox, \emph{{Lehninger principles of biochemistry}},
  fifth edition~ed.\hskip 1em plus 0.5em minus 0.4em\relax W. H. Freeman,
  January 2008.

\bibitem{Alpha2013Helix}
K.~Server, ``{Protein alpha helix},'' last accessed on November 26, 2013,
  [{knowledgeserver.wordpress.com}].

\bibitem{Khan2011BioTech}
F.~A. Khan, \emph{Biotechnology Fundamentals}.\hskip 1em plus 0.5em minus
  0.4em\relax CRC Press, Taylor \& Francis Group, 2011.

\bibitem{PDB2013Intro}
P.~D. Bank, ``{An information portal to biological macromolecular
  structures},'' last access on October 12, 2013, [www.rcsb.org/pdb/].

\bibitem{PyMOL2013Intro}
{Schr\"odinger, LLC}, ``The {PyMOL} molecular graphics system, version~1.3r1,''
  August 2010.

\bibitem{Anfinsen1973GovernFolding}
C.~B. Anfinsen, ``{The principles that govern the folding of protein chains},''
  \emph{Science}, vol. 181, no. 4096, pp. 223--230, 1973.

\bibitem{Tanford1968Denaturation}
C.~Tanford, ``Protein denaturation,'' \emph{Advances in Protein Chemistry},
  vol.~23, pp. 121--182, 1968.

\bibitem{Crick1958CentralDogma}
F.~Crick, ``{On protein synthesis},'' \emph{Symp. Soc. Exp. Biol.}, vol. XII,
  pp. 139--163, 1958.

\bibitem{crick1970dogma}
------, ``{Central dogma of molecular biology},'' \emph{Nature}, vol. 227, no.
  5258, pp. 561--563, 1970.

\bibitem{Anfinsen1961Dogma}
C.~B. Anfinsen, E.~Haber, M.~Sela, and F.~H. White, ``{The kinetics of
  formation of native ribonuclease during oxidation of the reduced polypeptide
  chain.}'' \emph{National Academy of Sciences, USA}, vol.~47, pp. 1309--1314,
  Sep. 1961.

\bibitem{Berg2002Bonds}
J.~L.~T. Jeremy M~Berg and L.~Stryer, \emph{Biochemistry, 5th edition (Section
  1.3 -- Chemical Bonds in Biochemistry)}.\hskip 1em plus 0.5em minus
  0.4em\relax New York: W H Freeman, 2002.

\bibitem{Gold1987HydrogenBond}
V.~Gold, \emph{{Compendium of chemical terminology}}.\hskip 1em plus 0.5em
  minus 0.4em\relax International Union of Pure and Applied Chemistry, 1987.

\bibitem{Beijer1998HydrogenBond}
F.~H. Beijer, H.~Kooijman, A.~L. Spek, R.~P. Sijbesma, and E.~W. Meijer,
  ``{Self-complementarity achieved through quadruple hydrogen bonding},''
  \emph{Angewandte Chemie International Edition}, vol.~37, no. 1-2, pp. 75--78,
  1998.

\bibitem{Lecture2Interaction}
B.~Methods, ``{Lecture: Protein interactions leading to folding},'' last
  accessed on December 18, 2013,
  [{www.chembio.uoguelph.ca/educmat/phy456/456lec02.htm}].

\bibitem{Kyte1982Amino}
J.~Kyte and R.~F. Doolittle, ``{Amino acid scale: hydropathicity},'' \emph{J.
  Mol. Biol}, vol. 157, pp. 105--132, 1982.

\bibitem{Dill1985Globular}
K.~A. Dill, ``{Theory for the folding and stability of globular proteins},''
  \emph{Biochemistry}, vol.~24, no.~6, pp. 1501--1509, 1985.

\bibitem{Science2005MuchMore2Know}
{The Science Editorial}, ``{So much more to know},'' \emph{The Science}, vol.
  309, no. 5731, pp. 78--102, July 2005.

\bibitem{Seringhaus2007ChemistryNobel}
M.~Seringhaus and M.~Gerstein, ``{Chemistry Nobel rich in structure},''
  \emph{Science}, vol. 315, no. 5808, p.~40, 2007.

\bibitem{Horton1992Energy}
N.~Horton and M.~Lewis, ``Calculation of the free energy of association for
  protein complexes,'' \emph{Protein science : a publication of the Protein
  Society}, vol.~1, no.~1, pp. 169--181, 1992.

\bibitem{King2012Prion}
O.~D. King, A.~D. Gitler, and J.~Shorter, ``{The tip of the iceberg:
  RNA-binding proteins with prion-like domains in neurodegenerative disease},''
  \emph{Brain Research}, vol. 1462, pp. 61--80, 2012.

\bibitem{Smith2003NatureEd}
{Adam Smith}, ``{Protein misfolding},'' \emph{Nature Reviews Drug Discovery},
  vol. 426, no. 6968, pp. 78--102, December 2003.

\bibitem{Dobson2003Misfold}
C.~M. Dobson, ``{Protein folding and misfolding},'' \emph{Nature}, vol. 426,
  no. 6968, pp. 884--890, 2003.

\bibitem{Chiti2006Misfolding}
F.~Chiti and C.~M. Dobson, ``{Protein Misfolding, Functional Amyloid, and Human
  Disease},'' \emph{Annu Rev Biochem}, vol. 75(1), pp. 333--366, 2006.

\bibitem{Mathias2013Review}
{Jucker Mathias} and {Walker Lary C.}, ``{Self-propagation of pathogenic
  protein aggregates in neurodegenerative diseases},'' \emph{Nature}, vol. 501,
  no. 7465, pp. 45--51, 2013.

\bibitem{Pandi1997Drug}
P.~Veerapandian, Ed., \emph{{Structure-based drug design}}.\hskip 1em plus
  0.5em minus 0.4em\relax New York: Marcel Dekker, 1997.

\bibitem{Breda2008Drug}
A.~Breda, N.~F. Valadares, O.~N. de~Souza, and R.~C. Garratt., ``{Protein
  Structure, Modelling and Applications},'' 2008.

\bibitem{Kast1997Engg}
P.~Kast and D.~Hilvert, ``{3D structural information as a guide to protein
  engineering using genetic selection},'' \emph{Current Opinion in Structural
  Biology}, vol.~7, no.~4, p. 470–479, 1997.

\bibitem{Berry2001Engg}
S.~M. Berry and Y.~Lu, \emph{Protein Structure Design and Engineering}.\hskip
  1em plus 0.5em minus 0.4em\relax John Wiley \& Sons, Ltd, 2001.

\bibitem{Lilie2003Biotech}
H.~Lilie, ``{Designer proteins in biotechnology},'' \emph{EMBO Reports},
  vol.~4, no.~4, pp. 346--351, 2006.

\bibitem{Rainer2003Biotech}
R.~Jaenicke and R.~Sterner, ``{Protein Design at the Crossroads of
  Biotechnology, Chemistry, Theory, and Evolution},'' \emph{Angewandte Chemie
  International Edition}, vol.~42, no.~2, pp. 140--142, 2003.

\bibitem{Dill2007FoldingProblem}
K.~A. Dill, S.~B. Ozkan, T.~R. Weikl, J.~D. Chodera, and V.~A. Voelz, ``{The
  protein folding problem: when will it be solved?}'' \emph{Current opinion in
  structural biology}, vol.~17, no.~3, pp. 342--346, 2007.

\bibitem{Dill2008FoldingProblemReview}
K.~A. Dill, S.~B. Ozkan, M.~S. Shell, and T.~R. Weikl, ``{The protein folding
  problem},'' \emph{Annual review of biophysics}, vol.~37, p. 289, 2008.

\bibitem{Lodish2012}
C.~A. K. M. K. A. B. H. P. A.~A. Harvey~Lodish, Arnold~Berk and M.~P. Scott,
  \emph{Molecular Cell Biology}.\hskip 1em plus 0.5em minus 0.4em\relax WH
  Freeman \& Co., New York, 2012.

\bibitem{Kurt2001NMR}
K.~Wuthrich, ``The way to nmr structures of proteins,'' \emph{Nature Struct Mol
  Biol}, vol.~8, no.~11, pp. 923 -- 925, 2001.

\bibitem{CryoEM2013}
J.~L.~S. Milne, M.~J. Borgnia, A.~Bartesaghi, E.~E.~H. Tran, L.~A. Earl, D.~M.
  Schauder, J.~Lengyel, J.~Pierson, A.~Patwardhan, and S.~Subramaniam,
  ``{Cryo-electron microscopy – a primer for the non-microscopist},''
  \emph{FEBS Journal}, vol. 280, no.~1, p. 28–45, 2013.

\bibitem{Bartesaghi2012CryoEM}
\BIBentryALTinterwordspacing
A.~Bartesaghi, F.~Lecumberry, G.~Sapiro, and S.~Subramaniam, ``{Protein
  Secondary Structure Determination by Constrained Single-Particle
  Cryo-Electron Tomography},'' \emph{Structure}, vol.~20, no.~12, p.
  2003–2013, 2012. [Online]. Available:
  \url{http://www.sciencedirect.com/science/article/pii/S0969212612004121}
\BIBentrySTDinterwordspacing

\bibitem{Rupp2004Crystallization}
B.~Rupp and J.~Wang, ``Predictive models for protein crystallization,''
  \emph{Methods}, vol.~34, no.~3, pp. 390 -- 407, 2004, macromolecular
  Crystallization.

\bibitem{Zhang2005PDB}
Y.~Zhang and J.~Skolnick, ``{The protein structure prediction problem could be
  solved using the current PDB library},'' \emph{Proceedings of the National
  Academy of Sciences of the United States of America}, vol. 102, no.~4, p.
  1029, 2005.

\bibitem{Bowie1991Threading}
J.~U. Bowie, R.~Luthy, and D.~Eisenberg, ``{A method to identify protein
  sequences that fold into a known three-dimensional structure},''
  \emph{Science}, vol. 253, no. 5016, p. 164, 1991.

\bibitem{Torda2005Threading}
A.~E. Torda, ``{Protein threading},'' \emph{The Proteomics Protocols Handbook},
  pp. 921--938, 2005.

\bibitem{Simons1999CASP3}
K.~T. Simons, R.~Bonneau, I.~Ruczinski, and D.~Baker, ``{\emph{Ab initio}
  protein structure prediction of {CASP III} targets using ROSETTA},''
  \emph{PROTEINS: Structure, Function, and Bioinformatics}, vol. Suppliment 3,
  pp. 171--176, 1999.

\bibitem{Baker2001StructuralGenomics}
D.~Baker and A.~Sali, ``{protein structure prediction and Structural
  Genomics},'' \emph{Science}, vol. 294, no. 5540, pp. 93--96, 2001.

\bibitem{Dill1997Levinthal}
K.~A. Dill and H.~S. Chan, ``{From Levinthal to pathways to funnels},''
  \emph{Nature Structural Biology}, vol.~4, no.~1, pp. 10--19, 1997.

\bibitem{Levinthal1968Pathway}
C.~Levinthal, ``{Are there pathways for protein folding?}'' \emph{Journal of
  Medical Physics}, vol.~65, no.~1, pp. 44--45, 1968.

\bibitem{Lesh2005Heuristic}
N.~Lesh, J.~Marks, A.~McMahon, and M.~Mitzenmacher, ``{New heuristic and
  interactive approaches to {2D} rectangular strip packing},'' \emph{Journal of
  Experimental Algorithmics (JEA)}, vol.~10, pp. 1--2, 2005.

\bibitem{Palu2004Constraint}
A.~{Dal Pal{\`u}}, A.~Dovier, and F.~Fogolari, ``{Constraint logic programming
  approach to protein structure prediction},'' \emph{BMC bioinformatics},
  vol.~5, no.~1, p. 186, 2004.

\bibitem{Backofen2006FastExact}
R.~Backofen and S.~Will, ``{A Constraint-based approach to fast and exact
  structure prediction in three-dimensional protein models},'' \emph{Constraint
  : Springer}, vol.~11, no.~1, pp. 5--30, 2006.

\bibitem{Unger1993GASimulation}
R.~Unger and J.~Moult, ``{Genetic algorithms for protein folding
  simulations},'' \emph{J. of Mol. Biology}, vol. 231, no.~1, pp. 75--81, 1993.

\bibitem{Hoque2007PhDThesis}
M.~T. Hoque, ``{Genetic Algorithm for {\emph{ab initio}} protein structure
  prediction based on low resolution models},'' Ph.D. dissertation, Monash
  University, Victoria, Australia, Sep. 2007.

\bibitem{Higgs2010Resampling}
T.~Higgs, B.~Stantic, M.~T. Hoque, and A.~Sattar, ``{Genetic algorithm
  feature-based resampling for protein structure prediction},'' pp. 1--8, 2010.

\bibitem{Backofen1999ExtAlphabet}
R.~Backofen, S.~Will, and E.~Bornberg-Bauer, ``{Application of constraint
  programming techniques for structure prediction of lattice proteins with
  extended alphabets},'' \emph{BIOINFORMATICS}, vol.~15, no.~3, pp. 234--242,
  1999.

\bibitem{Dill1995ExactModel}
K.~A. Dill, S.~Bromberg, K.~Yue, C.~H. Sun, K.~M. Ftebig, D.~P. Yee, and P.~D.
  Thomas, ``{Principles of protein folding -- a perspective from simple exact
  models},'' \emph{Protein Science}, vol.~4, pp. 561--602, 1995.

\bibitem{Hoque2009ExtendedHP}
M.~T. Hoque, M.~Chetty, and A.~Sattar, ``{Extended {HP} model for protein
  structure prediction},'' \emph{J. of Mol. Biology}, vol.~16, no.~1, pp.
  83--103, 2009.

\bibitem{Baker2006Macromolecular}
D.~Baker, ``{Prediction and design of macromolecular structures and
  interactions},'' \emph{Philosophical Transactions of the Royal Society B:
  Biological Sciences}, vol. 361, no. 1467, p. 459, 2006.

\bibitem{Ramachandran1963Chain}
G.~N. Ramachandran, C.~Ramakrishnan, and V.~Sasisekharan, ``{Stereochemistry of
  polypeptide chain configurations},'' \emph{J. of Mol. Biology}, vol.~7, no.
  1963, pp. 95--99, 1963.

\bibitem{Samudrala1999LowResolution}
R.~Samudrala, Y.~Xia, M.~Levitt, and E.~S. Huang, ``{A combined approach for
  \emph{ab-initio} construction of low resolution protein tertiary structures
  from sequence},'' vol.~4, pp. 505--516, 1999.

\bibitem{Hoque2009Review}
M.~T. Hoque, M.~Chetty, and A.~Sattar, ``{Genetic algorithm in {\emph{Ab
  Initio}} protein structure prediction using low resolution model: a
  review},'' \emph{Springer-Verlag Berlin Heidelberg}, vol. Biomedical data and
  application, no. 224, pp. 317--242, 2009.

\bibitem{Chivian2003Ab}
D.~Chivian, T.~Robertson, R.~Bonneau, and D.~Baker, ``{{\emph{Ab initio}}
  methods},'' 2003.

\bibitem{Bonneau2001Progress}
R.~Bonneau and D.~Baker, ``{{\emph{Ab initio}} protein structure prediction:
  progress and prospects},'' \emph{Annual Review of Biophysics and Biomolecular
  Structure}, vol.~30, no.~1, pp. 173--189, 2001.

\bibitem{Hoque2007FCC3D}
M.~T. Hoque, M.~Chetty, and A.~Sattar, ``{Protein folding prediction in {3D FCC
  HP} lattice model using genetic algorithm},'' vol. 2007.\hskip 1em plus 0.5em
  minus 0.4em\relax IEEE Congress on Evolutionary Computation, 2007, pp.
  4138--4145.

\bibitem{Hales2005Proof}
T.~C. Hales, ``{A proof of the Kepler conjecture},'' \emph{The Annals of
  Mathematics}, vol. 162, no.~3, pp. 1065--1185, 2005.

\bibitem{PetkovM09LBM}
K.~Petkov, F.~Qiu, Z.~Fan, A.~E. Kaufman, and K.~Mueller, ``{Efficient {LBM}
  visual simulation on face-centered cubic lattices},'' \emph{IEEE Trans. Vis.
  Comput. Graph.}, vol.~15, no.~5, p. 802–814, 2009.

\bibitem{Unger1992GA}
R.~Unger and J.~Moult, \emph{{Why genetic algorithms are suitable for protein
  folding analysis: The theoretical foundations}}.\hskip 1em plus 0.5em minus
  0.4em\relax University of Maryland, 1992.

\bibitem{Unger1993GAApp}
------, ``{On the applicability of genetic algorithms to protein folding},'' in
  \emph{{System Sciences, 1993, Proceeding of the Twenty-Sixth Hawaii
  International Conference on}}, vol.~1.\hskip 1em plus 0.5em minus 0.4em\relax
  IEEE, 1993, pp. 715--725.

\bibitem{Chen2006OffLattice}
M.~Chen and W.-q. Huang, ``{Heuristic algorithm for off-lattice protein folding
  problem},'' \emph{Journal of Zhejiang University-Science B}, vol.~7, no.~1,
  pp. 7--12, 2006.

\bibitem{Hart1997OffLattice}
W.~E. Hart and S.~Istrail, ``{Lattice and off-lattice side chain models of
  protein folding: linear time structure prediction better than 86\% of
  optimal},'' \emph{Journal of Computational Biology}, vol.~4, no.~3, pp.
  241--259, 1997.

\bibitem{Stillinger1993ToyAB}
F.~H. Stillinger, T.~Head-Gordon, and C.~L. Hirshfeld, ``{Toy model for protein
  folding},'' \emph{Physical Review E}, vol.~48, no.~2, p. 1469, 1993.

\bibitem{Wang2009ABModel}
T.~Wang and X.~Zhang, ``{3D Protein structure prediction with genetic tabu
  search algorithm in Off-Lattice AB model},'' vol.~1, pp. 43--46, 2009.

\bibitem{Berrera2003T20}
M.~Berrera, H.~Molinari, and F.~Fogolari, ``{Amino acid empirical contact
  energy definitions for fold recognition in the space of contact maps},''
  \emph{BMC Bioinformatics}, vol.~4, no.~1, p.~8, 2003.

\bibitem{Lazaridis2000EffEnergy}
T.~Lazaridis and M.~Karplus, ``{Effective energy functions for protein
  structure prediction},'' \emph{Current Opinion in Structural Biology},
  vol.~10, no.~2, pp. 139--145, 2000.

\bibitem{Bornberg1997Chain}
E.~Bornberg-Bauer, ``{Chain growth algorithms for HP-type lattice
  proteins}.''\hskip 1em plus 0.5em minus 0.4em\relax Research in Computational
  Molecular Biology, 1997.

\bibitem{Crippen1991Matrix}
G.~M. Crippen, ``{Prediction of protein folding from amino acid sequence over
  discrete conformation spaces},'' \emph{Biochemistry}, vol.~30, no.~17, pp.
  4232--4237, 1991.

\bibitem{Li1996Emergence}
H.~Li, R.~Helling, C.~Tang, and N.~Wingreen, ``{Emergence of preferred
  structures in a simple model of protein folding},'' \emph{Science}, vol. 273,
  no. 5275, p. 666, 1996.

\bibitem{Buchler2000Surveying}
N.~E.~G. Buchler and R.~A. Goldstein, ``{Surveying determinants of protein
  structure designability across different energy models and amino-acid
  alphabets: A consensus},'' \emph{The Journal of Chemical Physics}, vol. 112,
  p. 2533, 2000.

\bibitem{Miyazawa1985T20}
S.~Miyazawa and R.~L. Jernigan, ``{Estimation of effective interresidue contact
  energies from protein crystal structures: quasi-chemical approximation},''
  \emph{Macromolecules}, vol.~18, no.~3, pp. 534--552, 1985.

\end{thebibliography}
\end{small}

\end{document}